\documentclass[11pt, titlepage]{article}

\usepackage{enumitem} 
\usepackage{natbib} 
\usepackage{graphicx} 
\usepackage[ruled,vlined]{algorithm2e} 
\usepackage{amsmath,amssymb} 
\usepackage[margin=1in]{geometry}
\usepackage{setspace}
\usepackage{float} 
\usepackage[table]{xcolor} 
\usepackage{boldline} 
\usepackage{url} 

\defcitealias{PCAST2016}{PCAST, 2016}
\defcitealias{NIST2012}{NIST, 2012}
\defcitealias{NAS:2009}{NRC, 2009}

\definecolor{lightgray}{gray}{0.75}
\definecolor{lightergray}{gray}{0.9}

\usepackage{draftwatermark}
\SetWatermarkText{\textcopyright \ 2019 SDSU}
\SetWatermarkScale{3}

\begin{document}

\title{A Bayesian approach for the analysis of error rate studies in forensic science}
\author{J. H. Hendricks and C. Neumann}
\date{{Department of Mathematics and Statistics\\ South Dakota State University\\ AME Building\\ Box: 2225\\ Brookings, SD, USA.}\\ \hspace{10mm} \\{E-mail: jessiehhendricks@gmail.com, cedric.neumann@me.com}
}
\maketitle

\begin{abstract}
Over the past decade, the field of forensic science has received recommendations from the National Research Council of the U.S. National Academy of Sciences, the U.S. National Institute of Standards and Technology, and the U.S. President's Council of Advisors on Science and Technology to study the validity and reliability of forensic analyses. More specifically, these committees recommend estimation of the rates of occurrence of erroneous conclusions drawn from forensic analyses. ``Black box" studies for the various subjective feature-based comparison methods are intended for this purpose.

In general, ``black box" studies often have unbalanced designs, comparisons that are not independent, and missing data. These aspects pose difficulty in the analysis of the results and are often ignored. Instead, interpretation of the data relies on methods that assume independence between observations and a balanced experiment. Furthermore, all of these projects are interpreted within the frequentist framework and result in point estimates associated with confidence intervals that are confusing to communicate and understand.

We propose to use an existing likelihood-free Bayesian inference method, called Approximate Bayesian Computation (ABC), that is capable of handling unbalanced designs, dependencies among the observations, and missing data. ABC allows for studying the parameters of interest without recourse to incoherent and misleading measures of uncertainty such as confidence intervals. By taking into account information from all decision categories for a given examiner and information from the population of examiners, our method also allows for quantifying the risk of error for the given examiner, even when no error has been recorded for that examiner. This opens the door to the detection of behavioural patterns in the decision-making of examiners through their ABC rate estimates. These patterns could be used to detect error prone examiners, enabling additional training efforts to be more tailored to each examiner, limiting the risk of errors before they occur.

We illustrate our proposed method by reanalysing the results of the ``Noblis Black Box" study by Ulery et al. in 2011. We did not choose this study because we disagree with their results, but because it is a good example of a study with dependent observations and missing data, and the data is publicly available. The ABC estimates for the population generally agreed Ulery et al.'s plug-in estimates. However, credible intervals obtained from ABC are much wider than the confidence intervals for the corresponding parameter estimates that did not account for the dependencies among observations.
\end{abstract}

\spacing{1.5}

\section{Introduction}

Conclusions resulting from the examinations of forensic evidence are not exempt from errors. For example, following the ACE-V process, a fingerprint examiner may conclude that a fingermark does not originate from the same finger as a control impression, when it truly does.

Forensic analyses have received, and continue to receive, criticism \citep{Cole:2005, Zabell:2005, SaksKoehler:2008} due in part to erroneous conclusions that occur at an unknown (but potentially low) rate. Building on this criticism, a committee appointed by the National Academy of Sciences \citepalias{NAS:2009} recommended the study of the validity and reliability of forensic analyses, which includes the design and performance of scientific studies to estimate the rates of error present in the conclusions drawn from forensic analyses. 
 
Additional reports have also focused on the study of the rate of occurrence of erroneous conclusions. The National Institute of Standards and Technology (NIST) emphasised the importance of quantifying the rate of erroneous conclusions resulting from forensic analyses \citepalias{NIST2012}. They assert that knowledge about error rates enables the assessment of the reliability and allows for appropriate confidence to be placed on a given evidence. 


In 2016, the President's Council of Advisors on Science and Technology (PCAST) \citepalias{PCAST2016} emphasised that the validity of subjective feature-based comparison methods (such as fingerprint examination, firearms analysis, or footwear analysis) must be assessed through ``black box" studies that analyse the rates of occurrence of erroneous conclusions. They found that only several ``black box" studies have been performed for the evidence types they reviewed: the Noblis study by \cite{Ulery2011} and the Miami-Dade study by \cite{Pacheco2014}, both for fingerprint examination, and a study by \cite{Baldwin2014} for firearms analysis. PCAST recommended that additional ``black box" studies be conducted for the various subjective feature-based comparison methods to further study their validity and reliability. 

Whether we agree with the appropriateness for forensic scientists to report categorical conclusions, we can expect that additional error rate studies will be performed in the future to answer the recommendations from the NRC, NIST and PCAST. In any case, it is critical that the results of these studies are correctly analysed. However, complex experimental designs can make this difficult. For example, PCAST correctly points out that what they name ``set-based analyses'' result in dependent observations \citepalias{PCAST2016}. Estimating error rates based on such dependent observations cannot rely on straightforward plug-in estimates. However, PCAST incorrectly claims that the observations resulting from ``black box'' studies are independent. Firstly, from the design of ``black box'' studies, it is not clear whether the forensic scientists or the test cases are the experimental units. Secondly, forensic scientists and test cases are re-used throughout a given experiment, usually not in a balanced manner. This creates dependencies in the resulting observations that cannot be studied using the traditional tools developed in the context of the statistical design of experiments \citep{Casella2008}. 

In this paper, we propose a method to analyse the results of ``black box'' experiments that can handle the most complex designs in a rigorous manner. Our method is based on a class of computational methods called Approximate Bayesian Computation (ABC), which relies on strong statistical foundations with well-understood asymptotic properties. Our method can deal with unbalanced designs, dependent observations and missing data. Our method allows for studying the parameters of interest without recourse to incoherent and misleading measures of uncertainty such as confidence intervals. Finally, our method allows us to go beyond simply reporting point estimates for error rates: our method can result in predictive models that quantify the risk of errors of any given examiner, even when no error has been recorded for that examiner.

In this paper, we provide an example of the application of our method by reanalysing the results of the well-known ``Noblis Black Box'' study \citep{Ulery2011}. We have not chosen this study because we disagree with their results, but because it is a good example of a design that involves numerous uncontrollable factors, dependent observations, and missing data. Additionally, the original data collected during the study is publicly available \citep{BBResults}. 

Section \ref{BlackBox} summarises the study by \cite{Ulery2011} and provides a motivation for our approach. In section \ref{ABC}, we briefly discuss frequentist approaches for analysis of error rate studies, and then introduce a likelihood-free Bayesian approach (ABC) that can account for the dependencies among observations and missing data. In section \ref{Application}, we present the application of ABC to the data from the Ulery et al.'s study. In section \ref{Results}, we use our results to infer population and individual examiners' rates of erroneous conclusions resulting from the fingerprint examination process, and examine behavioural patterns in the decision making of fingerprint examiners. Finally, in section \ref{Conclusion}, we present our conclusions, and  recommendations for the analysis of future error rate studies and implementation of predictive models that may detect high-risk examiners.

\section{``Noblis Black Box'' study  \citep{Ulery2011}} 
\label{BlackBox}

Following the erroneous identification made by the FBI in connection with the 2004 terrorist attacks in Madrid, Spain, the review of the FBI's handling of the error \citep{OIG2006} and the research recommendations made by an FBI committee \citep{Budowle2006}, a ``black box'' study was performed by scientists at Noblis and reported in \cite{Ulery2011}. The purpose of this research project was to study the accuracy and reliability of decisions in forensic examination of fingerprints, and more specifically the rates at which fingerprint examiners make erroneous identifications or exclusions.  

During the first phase of the study, 169 participating examiners were each presented with a total of 100 pairs of fingermark and control impression out of 744 unique test cases prepared for the experiment. Out of the 16,900 possible conclusions, 16,873 were considered for the estimation of the error rates (27 were removed due to data entry error or inadvertent submission of multiple instances of the same test case to an examiner). During the second phase, 42 examiners from phase one were presented with several additional image pairs, resulting in an additional 248 responses considered for the data analysis. A total of 17,121 responses were considered to estimate the error rates reported by \citep{Ulery2011}.

For each pair of images, examiners were first asked to determine if the fingermark was of value for individualisation \footnote{We do not necessarily support the use of the term ``individualisation''; however, this was the term used in the original study.} (VID), value for exclusion (VEO), or no value (NV). If the image was deemed NV, the fingermark was not compared to the assigned control print. If the image was deemed either VID or VEO, the fingermark was compared with the control print. After the comparison stage, examiners were asked to provide a final decision of individualisation or exclusion, or to deem the comparison inconclusive. Therefore, a total of seven decision categories were possible: 
\spacing{1}
\begin{enumerate}
	\item No value (NV).
	\item Individualisation - value for individualisation (Ind. VID).
	\item Exclusion - value for individualisation (Exc. VID).
	\item Inconclusive - value for individualisation (Inc. VID).
	\item Individualisation - value for exclusion only (Ind. VEO).
	\item Exclusion - value for exclusion only (Exc. VEO).
	\item Inconclusive - value for exclusion only (Inc. VEO). 
\end{enumerate}
\spacing{1.5}
In the analysis of the experiment, these seven categories were further split into mated and non-mated scenarios, resulting in a total of 14 categories. The results obtained by \cite{Ulery2011} are reproduced in table \ref{tab:UleryS5} below.

\begin{table}[h]
\centering	
\resizebox{\textwidth}{!}{\begin{tabular}{|l|l|*{9}{r|}}
\hline
\multicolumn{1}{|c}{\cellcolor{lightgray}Comparison Decision} & \multicolumn{1}{c|}{\cellcolor{lightgray}Latent Value} & \multicolumn{1}{|c}{\cellcolor{lightgray}Total} & \multicolumn{1}{c}{\cellcolor{lightgray}Mates} & \multicolumn{1}{c|}{\cellcolor{lightgray}Non-mates} & \multicolumn{3}{|c|}{\cellcolor{lightgray}\% of mated pairs} & \multicolumn{3}{|c|}{\cellcolor{lightgray}\% of non-mated pairs} \\ 
\multicolumn{2}{|c|}{\cellcolor{lightgray}} & \multicolumn{3}{|c|}{\cellcolor{lightgray}} & \multicolumn{1}{|c}{\cellcolor{lightgray}PRES} & \multicolumn{1}{c}{\cellcolor{lightgray}CMP} & \multicolumn{1}{c|}{\cellcolor{lightgray}VID} & \multicolumn{1}{|c}{\cellcolor{lightgray}PRES} & \multicolumn{1}{c}{\cellcolor{lightgray}CMP} & \multicolumn{1}{c|}{\cellcolor{lightgray}VID} \\ \hlineB{2} 
\cellcolor{lightgray}(not compared) & \multicolumn{1}{l|}{\cellcolor{lightgray}NV} & \multicolumn{1}{|r|}{3,947} & 3,389 & \multicolumn{1}{r|}{558} & \multicolumn{1}{|r|}{29.3\%} & \multicolumn{1}{r|}{\cellcolor{lightergray}} & \multicolumn{1}{r|}{\cellcolor{lightergray}} & \multicolumn{1}{|r|}{10.1\%} & \multicolumn{1}{r|}{\cellcolor{lightergray}} & \multicolumn{1}{r|}{\cellcolor{lightergray}}\\ \hline
\cellcolor{lightgray}Exclusion & \multicolumn{1}{l|}{\cellcolor{lightgray}VEO} & \multicolumn{1}{|r|}{486} & 161 & \multicolumn{1}{r|}{325} & \multicolumn{1}{|r|}{1.4\%} & 2.0\% & \multicolumn{1}{r|}{\cellcolor{lightergray}} & \multicolumn{1}{|r|}{5.9\%} & 6.5\% & \multicolumn{1}{r|}{\cellcolor{lightergray}} \\ \hline
\cellcolor{lightgray}Exclusion & \multicolumn{1}{l|}{\cellcolor{lightgray}VID} & \multicolumn{1}{|r|}{4,072} & 450 & \multicolumn{1}{r|}{3,622} &  \multicolumn{1}{|r|}{3.9\%} & 5.5\% & \multicolumn{1}{r|}{7.5\%} & \multicolumn{1}{|r|}{65.3\%} & 72.7\% & 88.7\% \\ \hline
\cellcolor{lightgray}Inconclusive & \multicolumn{1}{l|}{\cellcolor{lightgray}VEO} & \multicolumn{1}{|r|}{2,596} & 2,019 & \multicolumn{1}{r|}{577} & \multicolumn{1}{|r|}{17.4\%} & 24.7\% & \multicolumn{1}{r|}{\cellcolor{lightergray}} & \multicolumn{1}{|r|}{10.4\%} & 11.6\% & \multicolumn{1}{r|}{\cellcolor{lightergray}} \\ \hline
\cellcolor{lightgray}Inconclusive & \multicolumn{1}{l|}{\cellcolor{lightgray}VID} & \multicolumn{1}{|r|}{2,311} & 1,856 & \multicolumn{1}{r|}{455} & \multicolumn{1}{|r|}{16.0\%} & 22.7\% & \multicolumn{1}{r|}{31.1\%} & \multicolumn{1}{|r|}{8.2\%} & 9.1\% & 11.1\% \\ \hline
\cellcolor{lightgray}Individualisation & \multicolumn{1}{l|}{\cellcolor{lightgray}VEO} & \multicolumn{1}{|r|}{40} & 40 & \multicolumn{1}{r|}{0} & \multicolumn{1}{|r|}{0.3\%} & 0.5\% & \multicolumn{1}{r|}{\cellcolor{lightergray}} & \multicolumn{1}{|r|}{0.0\%} & 0.0\% & \multicolumn{1}{r|}{\cellcolor{lightergray}} \\ \hline
\cellcolor{lightgray}Individualisation & \multicolumn{1}{l|}{\cellcolor{lightgray}VID} & \multicolumn{1}{|r|}{3,669} & 3,663 & \multicolumn{1}{r|}{6} & \multicolumn{1}{|r|}{31.6\%} & 44.7\% & \multicolumn{1}{r|}{61.4\%} & \multicolumn{1}{|r|}{0.1\%} & 0.1\% & 0.1\% \\ \hlineB{2}
\cellcolor{lightgray}Totals & \multicolumn{1}{l|}{\cellcolor{lightgray} } & \multicolumn{1}{|r|}{17,121} & 11,578 & \multicolumn{1}{r|}{5,543} & \multicolumn{1}{|r|}{100.0\%} & 100.0\% & \multicolumn{1}{r|}{100.0\%} & \multicolumn{1}{|r|}{100.0\%} & 100.0\% & 100.0\% \\ \hline
\cellcolor{lightgray}Total comparisons & \multicolumn{1}{l|}{\cellcolor{lightgray}Value (either)} & \multicolumn{1}{|r|}{13,174} & 8,189 & \multicolumn{1}{r|}{4,985} & \multicolumn{3}{|c|}{\cellcolor{lightergray}} & \multicolumn{3}{|c|}{\cellcolor{lightergray}} \\ \hline
\cellcolor{lightgray}Total comparisons & \multicolumn{1}{l|}{\cellcolor{lightgray}VID} & \multicolumn{1}{|r|}{10,052} & 5,969 & \multicolumn{1}{r|}{4,083} & \multicolumn{3}{|c|}{\cellcolor{lightergray}} & \multicolumn{3}{|c|}{\cellcolor{lightergray}} \\ \hline
\end{tabular}}
\caption{A reproduced version of table S5 from Ulery et al.'s Supplemental Information.}
\label{tab:UleryS5}
\end{table} 

This succinct description of the \citet{Ulery2011} experiment highlights many of the issues related to the analysis of most ``black box'' studies in general:
\spacing{1}
\begin{enumerate}
	\item The various comparisons are not independent:
	\begin{enumerate}
		\item The examination of several image pairs by each participant introduces dependencies between the resulting conclusions for that participant. 
		\item The examination of each test case by multiple participants introduces dependencies between the conclusions reached for a given test case. 
	\item Dependencies between participants may also be created by common training and operating procedures.
	\end{enumerate}
	\item The experiment is unbalanced:
	\begin{enumerate}
		\item Some participants examine more test cases than others.
		\item Examiners are not assigned an equal number of mated and non-mated image pairs, nor are they assigned the same number of mated and non-mated pairs as each other. 
		\item Some test cases are examined by more participants than others. Test cases are not assigned in a systematic manner.   
		\item Not all test cases result in the same number of decisions on the source of the fingermark since participants have the possibility to determine that a fingermark is not suitable for further examination.
	\end{enumerate}
	\item There are missing data. In the \cite{Ulery2011} study, some data were removed due to data entry error. Note that other studies, such as the one reported by \citet{Pacheco2014}, are more heavily affected by missing data than the one described above. For example, in the study by \citet{Pacheco2014}, many participants did not complete the experiment and only submitted results for a portion of the fingermarks they were asked to examine.
\end{enumerate}
\spacing{1.5}
Error rate studies, such as the ones considered by PCAST \citepalias{PCAST2016} and reported in \cite{Ulery2011} and \cite{Pacheco2014}, present point estimates, calculated using the plug-in principle, associated with confidence intervals. The dependencies in the design of the experiment, and the unbalanced and missing data make it inappropriate to use a binomial likelihood and, thus, to rely on plug-in estimates. In other words, not accounting for the various dependencies will result in underestimating the variance of the estimates, which will result in confidence/credible intervals that do not include the true error rates \footnote{Note that \cite{Ulery2011} recognised this point regarding their experiment but decided to move forward with confidence intervals based on the assumption of independent data (see section 2.1 in the appendix of \cite{Ulery2011}).}. We note that, when dependencies among the observations exist, u-statistics may provide more appropriate estimates \citep{Hoeffding1948}. 

In general, likelihood-based methods for the type of experiments designed to study error rates are impractical. For example, error rate data could be described by a generalised linear model with a fully specified covariance matrix. However, in the case of \cite{Ulery2011}'s study, this would require defining all dependencies in a 17,121 by 17,121 covariance matrix.


Traditionally, after using one of these methods to calculate parameter estimates, $(1-\alpha)100\%$ confidence intervals are usually constructed. As mentioned before, construction of a confidence interval should also account for any correlation present among the data. More importantly, interpretation of confidence intervals relies on ad aeternem theoretical repetition of the experiment and does not provide measures of uncertainty for the data at hand. Confidence intervals are not coherent measures of the uncertainty associated with the estimates, and are confusing and misleading even for statistical audiences. Upper bounds of confidence intervals are sometimes used to make dubious statements such as: ``the [false positive] rate could be as high as 1 in 18 cases" \citepalias{PCAST2016}. Indeed, the rate \textit{could} be as high as a certain value, but such statement does not inform us on the probability that this event would happen. 

It is worth mentioning that point estimates and confidence intervals do not allow for looking at the relationships between the different decision categories. A Bayesian approach for the study of the rates of decisions for the different categories enables to study the joint distributions of these parameters and make inference on how they affect each other, even in the presence of compositional data such as rates summing to 1.  In order to obtain reliable estimates of the rates of erroneous conclusions resulting from forensic examinations, it is of utmost importance to use statistical methods that can account for the dependencies that arise from complex experimental designs where the likelihood function is intractable. A class of likelihood-free inference methods called Approximate Bayesian Computation (ABC) provides a potential solution. 


\section{Approximate Bayesian Computation}
\label{ABC}

The desired result of an error rate analysis is an estimate, $\hat{\theta}$, of the rate(s) of interest, $\theta$, and a measure of uncertainty that displays information about the precision of the estimate. Additionally, in some cases, we may be interested in studying the effect of one or more factors (e.g., the difficulty of the comparison or the proficiency of the examiner) on the value of $\theta$. 

Bayesian methods offer an alternative approach for the analysis of error rate studies. Bayesian methods consider parameters as random variables that follow distributions. The distribution of a parameter can be used to discuss its value and assign the probability of any range of values for that parameter. These methods focus on an observed dataset, and do not rely on the hypothetical ad aeternem repetition of experiments mentioned above.  
Bayesian methods rely on the following proportionality relationship that can be produced using Bayes Theorem $$\pi(\theta|D_{obs}) \propto f(D_{obs}|\theta)\pi(\theta)$$ where $\pi(\theta|D_{obs})$ is the posterior distribution of the parameter $\theta$ given observed data, $D_{obs}$, $f(D_{obs}|\theta)$ is the likelihood function of $\theta$, given the observed data, and $\pi(\theta)$ is the prior density of $\theta$, containing all prior information on $\theta$. This result allows us to update prior belief (or knowledge) about the parameter $\theta$ using information contained in the data, $D_{obs}$, into a posterior belief. The posterior distribution quantifies the uncertainty about the value of the parameter. For example, using a posterior distribution, it is possible to state that there is a 97\% chance for the value of the parameter to be between two fixed values, $\theta_L$ and $\theta_U$, or that there is less than 0.1\% chance for the value of the parameter to be greater than an upper bound, $\theta_U$. Such interval is called a \textit{credible interval} and directly provides information on the likelihood of different values of $\theta$.

In some very simple scenarios, the form of the posterior distribution is a well-known parametric distribution (e.g., Normal, Dirichlet, Inverse Wishart) and parameter values for the posterior distribution can be derived easily. When the posterior distribution cannot be simplified to a known parametric distribution, sampling methods may be used instead to obtain a sample from the posterior distribution and study its properties (e.g., median, mean, credible interval). 

Unfortunately, standard Bayesian methods cannot be applied in all cases. Some scenarios require unreasonably complex models to describe the data so the corresponding likelihood function cannot be derived. This is the case for error rate experiments with complex dependency structures between conclusions. 

A class of methods, called Approximate Bayesian Computation (ABC) \citep{Beaumont2002}, allows for approximate Bayesian inference to be performed when the likelihood function is unavailable. The general idea of ABC is to replicate the experiment of interest by simulation using various parameter values. The pseudo-data resulting from these simulated experiments are then compared to the data acquired during the actual experiment. Parameter values that replicate the observed data well are retained, while parameter values that produce pseudo-data far from the observed data are discarded. The set of accepted parameter values is considered to be a sample from the approximate posterior distribution of the parameter of interest given the observed data.

ABC methods represent a shift in how experiments can be analysed. In a classic process, parameters are estimated from the data using a series of assumptions on how the data are generated. In the classic process, verifying the robustness of the assumptions may be difficult for datasets acquired using complex designs. In the ABC process, parameter values are proposed and pseudo-data are generated using the proposed values. The data generation process also relies on assumptions. However, should these assumptions be unrealistic, the resulting pseudo-data is unlikely to correspond to the observed data; thus, the assumptions are tested by the algorithm.

One of the standard ABC algorithms is an adaptation of the rejection sampling algorithm (see \cite{Sisson2019ch1}). It proceeds as follows: a candidate value for the parameter of interest, $\theta^{(i)}$, is sampled from a prior distribution, $\pi(\theta)$; that candidate value is used to generate pseudo-data, $D^{(i)}$, via a generative model (or algorithm), $\mathcal{M}$, that is defined in advance to reproduce the experiment or data generating process as best as possible; the candidate value, $\theta^{(i)}$, is kept if the pseudo-data is sufficiently similar to the observed data. The set of accepted values is considered to be a sample from the approximate posterior distribution of the parameter of interest. A standard version of an ABC rejection sampling algorithm, modified from \citet{Sisson2019ch1}, is presented in algorithm \ref{alg:ABCalgorithm}. 
\begin{center}
\scalebox{0.95}{\begin{algorithm}[H]         
    \DontPrintSemicolon
    \SetAlgoLined
	\SetKwInOut{Input}{Input}
	\SetKwInOut{Output}{Output}
	\Input{observed data $D_{obs}$, generative model $\mathcal{M}$, summary statistic $S(\cdot)$, distance metric $\Delta(\cdot)$, threshold $h$, and $N \in \mathbb{Z}^+$.}
    \For{i = 1 to N}{
			$\cdot$ Sample $\theta^{(i)} \sim \pi(\theta)$.\\
			$\cdot$ Generate data, $D^{(i)}$, from $\mathcal{M}$ given $\theta^{(i)}$.\\
			$\cdot$ summarise $D^{(i)}$ as $S(D^{(i)})=s^{(i)}$.\\
			$\cdot$ If $\Delta(s_{obs},s^{(i)}) \le h$ , accept $\theta^{(i)}$. Else return to step 1.		        
    }
 	\caption{ABC rejection sampler (modified from \cite{Sisson2019ch1})}
 	\Output{A sample $\theta^{(1)}$, ..., $\theta^{(N)}$ from $\hat{\pi}(\theta|s_{obs})$.} 
 	\label{alg:ABCalgorithm}
\end{algorithm}}
\end{center}

\newpage
In the algorithm, the level of similarity between the pseudo-data and the observed data is assessed using a distance metric and a threshold. To avoid the curse of dimensionality resulting from comparing high-dimensional vectors, the algorithm relies on summary statistic functions to represent the data in a much lower dimension. Reducing the dimension can increase the chance of producing pseudo-data that is considered similar to the observed data, leading to a better approximation. The summary statistic should ideally be sufficient to prevent information loss. In settings where a sufficient statistic is not available (which are most application settings), additional considerations must be made. For example, there is a tradeoff between the loss of information due to an insufficient summary statistic and the curse of dimensionality. The goal is to minimise the loss of information while also minimising the dimension of the vector \citep{Sisson2019ch5}. 

Inexact matching of the summarised pseudo-data and the summarised observed data is allowed to improve the efficiency of the algorithm by increasing the number of candidate parameter values that are accepted and then used to compose the sample from the approximate posterior distribution of the parameter of interest.
 
Algorithms (such as the one presented above) that accept $\theta^{(i)}$ for which $\Delta(s_{obs},s^{(i)})$ is far from 0 allow bias to be introduced into the approximation. While one may consider lowering the tolerance, $h$, towards 0 to reduce the bias, this causes the majority of samples to be rejected and results in a very inefficient sampling algorithm. An alternative approach involves performing an ``adjustment'' to the sampled set of $\theta^{(i)}$ to reduce the bias \citep{Beaumont2002, Blum2010, Sisson2019ch3}. The sampling steps of the algorithm remain unchanged, but additional steps for the adjustment are added at the end. Algorithm \ref{alg:ABCadjustment} presents an ABC rejection sampler with one such adjustment step \footnote{This particular adjustment step assumes heteroscedasticity of the error terms. Other adjustments are possible.}.  

\begin{center}
\scalebox{0.95}{\begin{algorithm}[H]         
    \DontPrintSemicolon
    \SetAlgoLined
	\SetKwInOut{Input}{Input}
	\SetKwInOut{Output}{Output}
	\Input{observed data $D_{obs}$, generative model $\mathcal{M}$, summary statistic $S(\cdot)$, distance metric $\Delta(\cdot)$, and number of simulations $N$.}
    \For{i = 1 to N}{
			$\cdot$ Sample $\theta^{(i)} \sim \pi(\theta)$.\\
			$\cdot$ Generate data, $D^{(i)}$, from $\mathcal{M}$ given $\theta^{(i)}$.\\
			$\cdot$ summarise $D^{(i)}$ as $S(D^{(i)})=s^{(i)}$. 					        
    }
    \ Obtain adjusted parameters as $\tilde{\theta}^{(i)} = \hat{m}(s_{obs}) + \dfrac{\hat{\sigma}(s_{obs})}{\hat{\sigma}(s^{(i)})}\left(\theta^{(i)} - \hat{m}(s^{(i)})\right)$, where $\hat{m}(\cdot)$ is a weighted regression model learned by regressing $\theta^{(i)}$ on $s^{(i)}$ for all $i$, and $\hat{\sigma}(\cdot)$ is also a weighted regression model, learned by regressing the set of squared residuals (from $\hat{m}(\cdot)$) on $s^{(i)}$ for all $i$. The weights depend on $\Delta(\cdot)$.
 	\caption{ABC rejection sampler with parameter adjustment}
 	\Output{An adjusted sample $\tilde{\theta}^{(1)}$, ..., $\tilde{\theta}^{(M)}$ from $\hat{\pi}(\theta|s_{obs})$, where $M \le N$.} 
 	\label{alg:ABCadjustment}
\end{algorithm}}
\end{center}


The sample $\tilde{\theta}^{(1)}$, ..., $\tilde{\theta}^{(M)}$ \footnote{It may be noted that the sample is drawn from $\hat{\pi}(\theta|s_{obs})$, the approximate posterior distribution of $\theta$, given the summary statistic of the observed data, $s_{obs}$. This is equivalent to $\hat{\pi}(\theta|D_{obs})$ when $S(\cdot)$ is a sufficient summary statistic.} can be used to assign a parameter estimate, $\hat{\theta}$, such as the mean, mode or median, and an empirical measure of the  probability of any range of values for~$\theta$. 

ABC has been used to analyze the results of experiments from a wide range of disciplines (see \cite{Sisson2019ch1} for an extensive list of applications) where the data may be unbalanced, high-dimensional, or encapsulate many different variable types. In the following sections, we illustrate the application of an ABC method to the experiment performed by \citet{Ulery2011}.

\section{Application of ABC to the ``Noblis Black Box'' study}
\label{Application}

The implementation of an ABC algorithm requires a data generating algorithm, $\mathcal{M}$, to replicate the experiment as closely as possible. We present a data generating algorithm for the ``Noblis Black Box" study in algorithm \ref{alg:BBdatagen}. The algorithm requires several inputs for examiners 1 through 169: 
\spacing{1}
\begin{enumerate}
	\item The number of mated image pairs, $n_{m}^{(j)}$, and the number of non-mated image pairs, $n_{nm}^{(j)}$, presented to examiner $j$. Values from $n_{m}^{(j)}$ and $n_{nm}^{(j)}$ are equal to the number of mated and non-mated pairs considered for analysis by \cite{Ulery2011} for each of their participants. 
	\item A vector of length seven, $\theta_{m}^{(j)}$, containing rates at which examiner $j$ classifies a mated image pair into one of the seven decision categories (NV, Ind. VID, Exc. VID, Inc. VID, Ind. VEO, Exc. VEO, and Inc. VEO). Each vector $\theta_{m}^{(j)}$ is sampled from a prior distribution over rates for the seven decision categories (see algorithm \ref{alg:ABCBlackBox}). 
	\item Another vector of length seven, $\theta_{nm}^{(j)}$, containing the rates at which examiner $j$ classifies a non-mated image pair into one of the seven decision categories. Each vector $\theta_{nm}^{(j)}$ is sampled from a prior distribution over rates for the seven decision categories (see algorithm \ref{alg:ABCBlackBox}). 
\end{enumerate}
\spacing{1.5}
	Inputs for the $j^{th}$ examiner are used to define the parameters of two multinomial distributions. The first multinomial distribution is used to draw a vector of counts, simulating the number of decisions in each category made by examiner $j$ for the $n_{m}^{(j)}$ mated image pairs assigned to this examiner. The second multinomial distribution is used to draw another vector of counts, simulating the decisions made by examiner $j$ for the $n_{nm}^{(j)}$ non-mated image pairs.  

\begin{center} 
	\begin{algorithm}[H]         
	    \DontPrintSemicolon
	    \SetAlgoLined
		\SetKwInOut{Input}{Input}
		\SetKwInOut{Output}{Output}
		\Input{$n_{m}^{(j)}$, $n_{nm}^{(j)}$, $\theta_{m}^{(j)}$, $\theta_{nm}^{(j)}$, for $j = 1,2, ..., 169$}
	    \For{j = 1 to 169}{
				$\cdot$ Sample a vector of decision counts, $x_{m}^{(j)}$, for the $n_{m}^{(j)}$ mated pairs presented to examiner $j$: $x_{m}^{(j)} \sim \text{Multinomial}(n_{m}^{(j)}, \theta_{m}^{(j)})$. \\
				$\cdot$ Sample a vector of decision counts, $x_{nm}^{(j)}$, for the $n_{nm}^{(j)}$ non-mated pairs presented to examiner $j$: $x_{nm}^{(j)} \sim \text{Multinomial}(n_{nm}^{(j)}, \theta_{nm}^{(j)})$. 
		}
	 	\caption{Data generating process for \citet{Ulery2011} experiment}
	 	\Output{Pseudo-decision counts for each of the 169 examiners.} 
	 	\label{alg:BBdatagen}
	\end{algorithm}
\end{center}

The data generating algorithm presented in algorithm \ref{alg:BBdatagen} accounts for the dependencies among decisions from a given examiner $j$ by fixing the rate vectors, $\theta_{m}^{(j)}$ and $\theta_{nm}^{(j)}$, to simulate all decisions made by that examiner during one iteration of the simulated experiment. However, the algorithm does consider that, for a given examiner, all test cases are independent and identically distributed $\text{Multinomial}(\theta_{\cdot}^{(j)})$.
Thus, in its current form, the data generating process does not account for the dependencies between the decisions made by multiple examiners on the same test case. It may be possible to imagine a more complex data generating algorithm that would also account for these dependencies. One such possibility would include defining an examiner's decision rate vectors as functions of a measure of the difficulty of a test case. This would result in decision rate vectors potentially being more strongly weighted towards inconclusive results for low quality fingermarks than good ones. This extension of the algorithm would require the use of data, such as the objective measures of image quality for each test case used by \citet{Ulery2014}, which are not currently publicly available. 

The ABC algorithm for the ``Noblis Black Box" experiment is presented in algorithm \ref{alg:ABCBlackBox}. 

\scalebox{1}{\begin{algorithm}[H]         
    \DontPrintSemicolon
    \SetAlgoLined
	\SetKwInOut{Input}{Input}
	\SetKwInOut{Output}{Output}
	\Input{number of simulations $N$, hyper-parameters for the population error rates $\alpha_{m}$, $\alpha_{nm}$, $\mu$ and $\sigma$, the numbers of mated and non-mated image pairs presented to examiner $j$, $n^{(j)}_{m}$ and $n^{(j)}_{nm}$, for $j=1,...,169$, observed data, $D_{obs}$, summary statistic $S(\cdot)$, and distance metric $\Delta(\cdot)$.}
    \For{i = 1 to N}{
			$\cdot$ Sample a vector of population rates, $\eta_{m}^{(i)} \sim \text{Dirichlet}(\alpha_m)$.\\
			$\cdot$ Sample a vector of population rates, $\eta_{nm}^{(i)} \sim \text{Dirichlet}(\alpha_{nm})$.\\
			$\cdot$ Sample a scale parameter, $\lambda^{(i)} \sim \text{logNormal}(\mu, \sigma^2)$.\\	
			\For{j = 1 to 169}{
				$\cdot$ Sample a vector of mated decision rates for examiner $j$, \\ \ $\theta_{m}^{(i,j)} \sim \text{Dirichlet}(\lambda^{(i)} \eta_{m}^{(i)})$.\\
				$\cdot$ Sample a vector of non-mated decision rates for examiner $j$, \ $\theta_{nm}^{(i,j)} \sim \text{Dirichlet}(\lambda^{(i)} \eta_{nm}^{(i)})$.\\
			}
			$\cdot$ Generate 17,121 total pseudo-decisions using the generative model in algorithm  \ref{alg:BBdatagen}, $D^{(i)} \sim \mathcal{M}|n_{m}^{(1)}, ..., n_{m}^{(169)}, n_{nm}^{(1)}, ..., n_{nm}^{(169)}, \theta_{m}^{(1,i)}, \theta_{m}^{(169,i)}, \theta_{nm}^{(1,i)}, \theta_{nm}^{(169,i)}$.\\
    }
	$\cdot$ summarise $D_{obs}$ as the total number of decisions in each category, $s_{obs}$.\\
	$\cdot$ summarise $D^{(i)}$ as the total number of decisions in each category, $s^{(i)}$, for all $i$.\\
    $\cdot$ Obtain adjusted parameters as $\tilde{\eta}_{(\cdot)}^{(i)} = \hat{m}(s_{obs}) + \dfrac{\hat{\sigma}(s_{obs})}{\hat{\sigma}(s^{(i)})}\left(\eta_{(\cdot)}^{(i)} - \hat{m}(s^{(i)})\right)$.\\ 
    \For{j = 1 to 169}{
		$\cdot$ summarise $D_{obs}$ as $s^{(j)}_{obs}$ and $D^{(i)}$ as $s^{(i,j)}$, the concatenation of the total number of decisions in each category for examiner $j$ and the total number of decisions in each category for the observed and the simulated experiment, separately. \\
    	$\cdot$ Obtain adjusted parameters as $\tilde{\theta}_{(\cdot)}^{(i,j)} = \hat{m}(s^{(j)}_{obs}) + \dfrac{\hat{\sigma}(s^{(j)}_{obs})}{\hat{\sigma}(s^{(i,j)})}\left(\theta_{(\cdot)}^{(i,j)} - \hat{m}(s^{(i,j)})\right)$. 
	}
 	\caption{ABC for the ``Noblis Black Box" study \citep{Ulery2011}}
 	\Output{Joint posterior samples, $\tilde{\eta}_{m}^{(1)},...,\tilde{\eta}_{m}^{(M)}$ and $\tilde{\eta}_{nm}^{(1)},...,\tilde{\eta}_{nm}^{(M)}$, for population rates and joint posterior samples for each examiner, $\tilde{\theta}_{m}^{(1,j)},...,\tilde{\theta}_{m}^{(M,j)}$ and $\tilde{\theta}_{nm}^{(1,j)},...,\tilde{\theta}_{nm}^{(M,j)}$, where $j=1,...,169$ and $M \le N$.} 
 	\label{alg:ABCBlackBox}
\end{algorithm}}

\newpage
In the algorithm, the ``Black Box" experiment is simulated $N$ times. In each simulation, we make several assumptions:
\spacing{1}
\begin{enumerate}
	\item Examiners are exchangeable.
	\item Test cases are exchangeable.
	\item Population rates can be represented by a single vector for mated test cases and a single rate vector for non-mated test cases.
	\item Dirichlet distributions with fixed hyperparameters propose plausible vectors of population rates.
	\item Rates for individual examiners can diverge from the population rates \footnote{This is one of the main points of contention in the interpretation of the point estimates in the ``Black Box" study. The point estimates are an estimate of the population average, but do not represent the variability of the population of examiners. Our solution provides an answer to this criticism.}.
	\item Dirichlet distributions with parameters based on the sampled population rates propose plausible vectors of individual examiner rates.
	\item The data generating algorithm reproduces the ``Black Box" experiment as closely as possible.
\end{enumerate}
\spacing{1.5}

Assumptions 1 and 2 result in rates, sampled under assumptions 4 and 6, that are independent and identically distributed. As mentioned before, it may be possible to constrain those vectors of rates and express them as functions of the difficulty of test cases and factors related to the proficiency of the examiners \footnote{Data, such as that collected by \citet{Ulery2014}, may allow these modifications to be made.}. 

At the conclusion of all $N$ simulated experiments, we obtain a joint posterior sample of rates for the population, as well as for each examiner. The adjustment process of the sample for population rates ``selects'' the rate values that result in pseudo-data that only differ slightly from the observed decision counts and adjust their values by weighted regression. This results in samples from the approximate posterior distributions, $\hat{\pi}(\eta_{m}|D_{obs})$ and $\hat{\pi}(\eta_{nm}|D_{obs})$. The adjustment process of the samples for each examiner ``selects'' the rate values that are most appropriate for each real examiner and ``adjusts'' their values in a similar manner as for the population rates. This results in samples from the approximate posterior distributions, $\hat{\pi}(\theta^{(j)}_{m}|D_{obs})$ and $\hat{\pi}(\theta^{(j)}_{nm}|D_{obs})$, for $j=1,...169$. Since the adjustment process ``selects'' the rate values that are most appropriate for each real examiner separately based on the specific observed counts for that examiner, examiners are no longer considered exchangeable after the simulations have been performed.  
 

Both algorithms \ref{alg:BBdatagen} and \ref{alg:ABCBlackBox} are implemented in R programming software. The \textit{abc} R package by \cite{Csillery2012} is used to perform a nonlinear heteroscedastic weighted regression adjustment using the ``neuralnet" option. 

To reanalyse the results of the ``Noblis Black Box" experiment, we use algorithm \ref{alg:ABCBlackBox} where the algorithm inputs are defined as follows:
\spacing{1}
\begin{enumerate}
	\item The number of simulations, $N$, is 1,000,000.
	\item The parameters of the Dirichlet distribution for the decision rates of the population for mated comparisons are $\alpha_{m}=[3, 0.5, 0.5, 2, 2, 0.5, 3]$. 
	\item The parameters of the Dirichlet distribution for the decision rates of the population for non-mated comparisons are $\alpha_{nm}=[1, 1, 7, 1, 1, 0.5, 0.5]$.
	\item The parameters for the lognormal distribution for a scale parameter used in sampling of the decision rates for the examiners from the population rates are $\mu=6$ and $\sigma=1$.
	\item The numbers of mated and non-mated image pairs $n^{(j)}_{m}$ and $n^{(j)}_{nm}$ presented to examiners $j=1,...169$ are obtained from the observed counts for each examiner in the ``Noblis Black Box" study.
	\item The observed data, $D_{obs}$, is the set of observed decisions from the ``Noblis Black Box" study.
\end{enumerate}
\spacing{1.5}

The values for the parameters of the Dirichlet distributions were selected so that their expectation would not be too far from the point estimates calculated by \cite{Ulery2011}. Other values are possible, although a much larger \textit{N} may be required to obtain a larger posterior sample of the parameters of interest. 

Note that all simulations and results from our analysis of the Noblis data account for the number of pairs of images that were presented to the examiners (``PRES'' columns in table \ref{tab:UleryS5}). It would be trivial to focus only on the pairs of images that have been compared (VID or VEO) or deemed of value for individualisation (VID) by adjusting the length of the vectors of rates, $\eta_{m}, \eta_{nm}, \theta_{m},\theta_{nm}$, and the number of pairs considered for each examiner, $n^{(j)}_{m}$, $n^{(j)}_{nm}$, in algorithms \ref{alg:BBdatagen} and \ref{alg:ABCBlackBox}.

\section{Results and discussion}
\label{Results}

\subsection{Population decision rates}
\label{PopulationResults}

Algorithm \ref{alg:ABCBlackBox} produces joint posterior samples of the 14 population decision rates based on the data collected in the ``Noblis Black Box" experiment. From this data, we define the point estimate for each of the 14 rates as the median of each marginal posterior sample. Since the ABC estimates are based on posterior samples of the joint distribution of all 14 rate parameters, they incorporate information on the relationships between the decision categories. Marginal posterior samples and ABC rate estimates for the 14 population decision rates are presented in figure \ref{fig:ResultsABCPopulation} below. We emphasise (as done by \cite{Ulery2011} in section 2.1 of their appendix) that our results are only valid for a population of examiners that is similar to the participants to the Noblis study, and which is performing similar tasks on fingerprint images with similar level of difficulty as those in the Noblis study.  

\begin{figure}[H]
	\centering
	\includegraphics[scale=0.5]{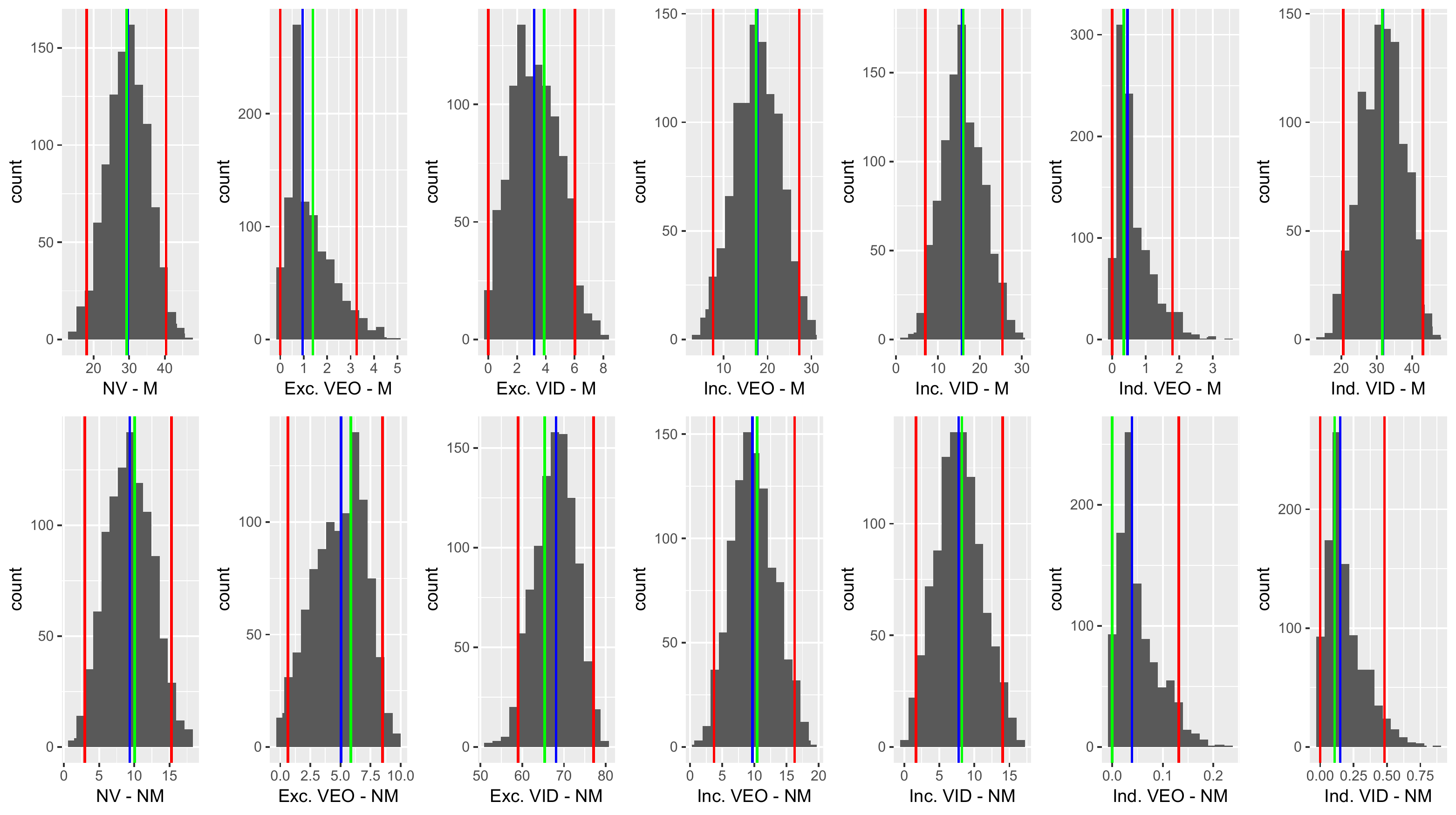}
	\caption{Histograms of the marginal posterior samples of rates for mated test cases are presented in the top row, while those for non-mated test cases are presented in the bottom row. Rates are labeled in percent form. Vertical blue lines indicate the ABC rate estimates, defined as the median of each marginal posterior sample. The vertical red lines indicate the lower and upper bounds of the 95\% highest posterior density interval for each marginal posterior sample. Vertical green lines indicate the \citet{Ulery2011} plug-in estimates for the rates of each decision category. Note that the plug-in estimates correspond to the ``PRES" columns in table \ref{tab:UleryS5}.}
	\label{fig:ResultsABCPopulation}
\end{figure}

In figure 1, we observe that ABC rate estimates and plug-in rate estimates are quite similar in most of the 14 decision categories. In fact, the vertical bars representing ABC rate estimates and plug-in rate estimates nearly overlap in several decision categories (NV-M, \ Inc.VEO-M, \ Inc.VID-M, \ Ind.VID-M). It follows that plug-in estimates also fall inside the 95\% highest posterior density intervals for most rates. The values of the plug-in estimate for the Ind.VEO-NM category and the lower bound of the 95\% highest posterior density interval are so close that the lines representing them overlap.

ABC population rate estimates and 95\% highest posterior density intervals for the 14 decision categories are presented in table \ref{tab:ResultsABCPopulation}.  

\begin{table}[H]
\centering
\resizebox{0.75\textwidth}{!}{\begin{tabular}{|l|c|c|c|c|}
  \hline
  \cellcolor{lightgray} & \cellcolor{lightgray}Mated est. & \cellcolor{lightgray}Mated HDI & \cellcolor{lightgray}Non-mated est. & \cellcolor{lightgray}Non-mated HDI \\ \hline
  \cellcolor{lightgray}NV & 29.61\% & [18.04\%, 40.36\%] & 9.35\% & [2.98\%, 15.30\%] \\ \hline
  \cellcolor{lightgray}Exc. VEO & 0.95\% & [0.00\%, 3.26\%] & 5.06\% & [0.63\%, 8.50\%] \\ \hline
  \cellcolor{lightgray}Exc. VID & 3.18\% & [0.00\%, 6.03\%] & 68.09\% & [58.98\%, 77.11\%] \\ \hline
  \cellcolor{lightgray}Inc. VEO & 17.67\% & [7.58\%, 27.25\%] & 9.71\% & [3.72\%, 16.28\%] \\ \hline
  \cellcolor{lightgray}Inc. VID & 15.69\% & [6.99\%, 25.35\%] & 7.75\% & [1.67\%, 14.02\%] \\ \hline
  \cellcolor{lightgray}Ind. VEO & 0.46\% & [0.00\%, 1.80\%] & 0.04\% & [0.00\%, 0.13\%] \\ \hline 
  \cellcolor{lightgray}Ind. VID & 31.81\% & [20.55\%, 43.02\%] & 0.15\% & [0.00\%, 0.48\%] \\
  \hline
\end{tabular}}
\caption{ABC population rate estimates and 95\% highest posterior density intervals for each of the 14 decisions. Values are presented as percents of total mated or non-mated presented pairs and are rounded to two decimal places. Note that since the medians are used as point estimates, they do not sum to 100\% across the seven categories for mated comparisons and the seven categories for non-mated comparisons.}
\label{tab:ResultsABCPopulation}
\end{table}

Several rates of interest for the population include: Exc. VEO and Exc. VID for mated test cases (false exclusion) and Ind. VEO and Ind. VID for non-mated test cases (false identification). From table \ref{tab:ResultsABCPopulation}, we see that: 
\spacing{1}
\begin{enumerate}
	\item An estimated 0.95\% of presented mated test cases result in an Exc. VEO conclusion, and there is a 95\% chance that the true rate falls between 0.00\% and 3.26\%. According to the Agresti-Coull method used by \cite{Ulery2011}, the corresponding confidence interval for the true rate of false exclusion for VEO fingermarks would range from 1.19\% to 1.62\%.  
	\item An estimated 3.18\% of presented mated test cases result in an Exc. VID conclusion, and there is a 95\% chance that the true rate falls between 0.00\% and 6.03\% vs. a confidence interval for the true rate of false exclusion for VID fingermarks ranging from 3.55\% to 4.25\%. 
	\item An estimated 0.04\% of presented non-mated test cases result in an Ind. VEO conclusion, and there is a 95\% chance that the true rate falls between 0.00\% and 0.13\% vs. a confidence interval for the true rate of false identification for VEO fingermarks ranging from -0.01\% to 0.08\%.
	\item An estimated 0.15\% of presented non-mated test cases result in an Ind. VID conclusion, and there is a 95\% chance that the true rate falls between 0.00\% and 0.48\% vs. a confidence interval for the true rate of false identification for VID fingermarks ranging from 0.04\% to 0.24\%. 
\end{enumerate}
\spacing{1.5}

The differences between the \cite{Ulery2011} plug-in estimates (in ``PRES" columns of table \ref{tab:UleryS5}) and the ABC rate estimates may be explained by the fact that the ABC estimates take into account some of the dependencies between conclusions on the image pairs and incorporate the relationships between the decision categories, while the plug-in estimates do not. In particular, we note that the credible intervals obtained using our method are much wider than the Agresti-Coull confidence intervals, that they are coherent (none of them suggest negative rates) and that they have an intuitive interpretation. That said, we note that, overall, the ABC rate estimates and the plug-in estimates from \cite{Ulery2011} agree in general.

\subsection{Examiner decision rates}
\label{ExaminerResults}

\begin{figure}[H]
	\centering
	\includegraphics[scale=0.4]{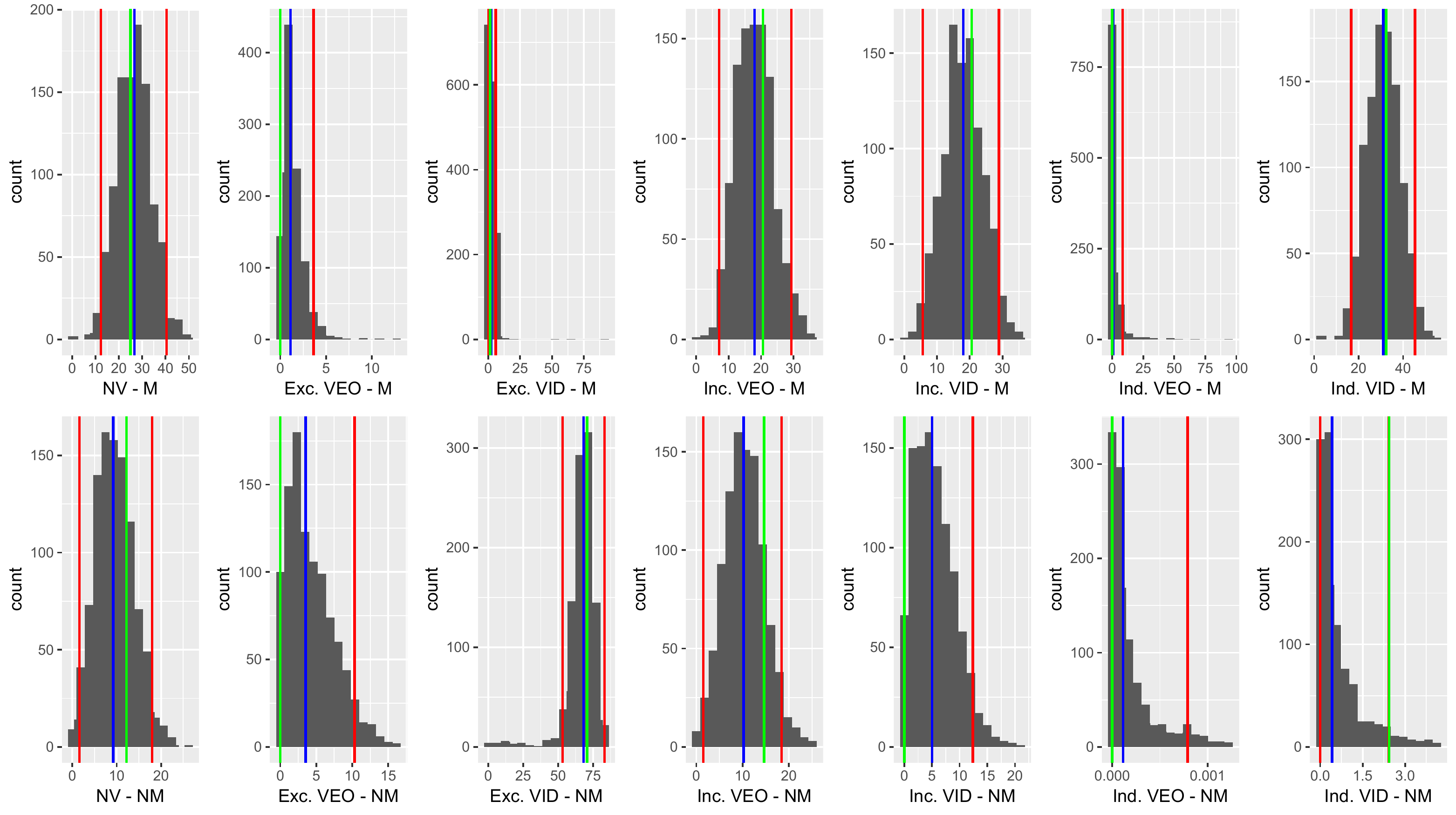}
	\caption{Histograms of the marginal posterior samples of rates (labeled in percent form) from examiner 28. Rates for mated test cases are presented in the top row, while those for non-mated test cases are presented in the bottom row. Vertical blue lines indicate the ABC rate estimates for examiner 28, defined as the median of each marginal posterior sample. The vertical red lines indicate the lower and upper bounds of a 95\% highest posterior density interval for each marginal posterior sample. Vertical green lines indicate the plug-in estimate for the rate of each decision category. Plug-in estimates of the rates for each examiner are obtained in a similar manner as the plug-in estimates for the population rates calculated by \citet{Ulery2011}.}
	\label{fig:ResultsABCExaminer28plot}
\end{figure}

\begin{figure}[H]
	\centering
	\includegraphics[scale=0.4]{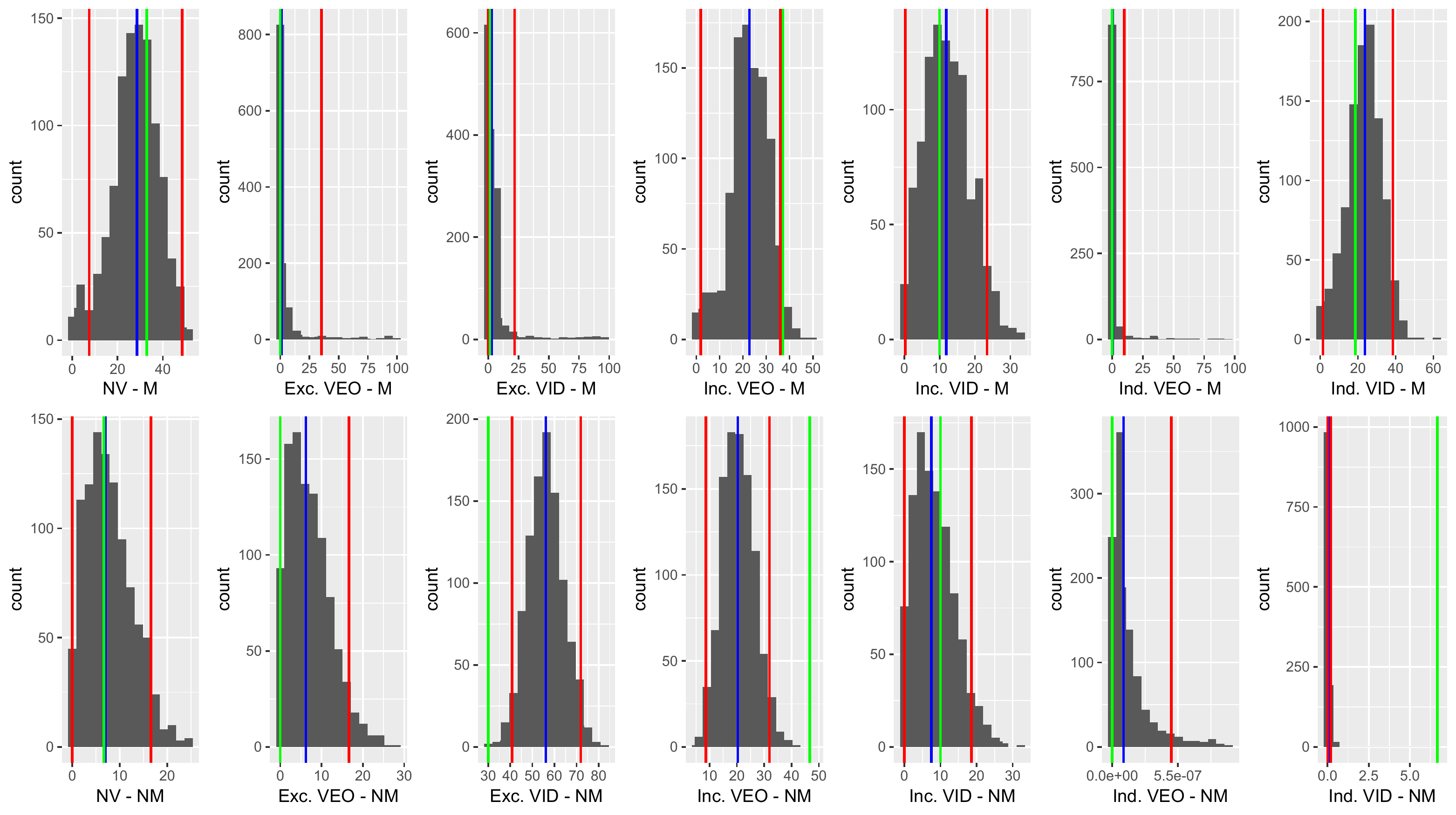}
	\caption{Histograms of the marginal posterior samples of rates (labeled in percent form) from examiner 63. Rates for mated test cases are presented in the top row, while those for non-mated test cases are presented in the bottom row. Vertical blue lines indicate the ABC rate estimates for examiner 63, defined as the median of each marginal posterior sample. The vertical red lines indicate the lower and upper bounds of a 95\% highest posterior density interval for each marginal posterior sample. Vertical green lines indicate the plug-in estimate for the rate of each decision category. Plug-in estimates of the rates for each examiner are obtained in a similar manner as the plug-in estimates for the population rates calculated by \citet{Ulery2011}.}
	\label{fig:ResultsABCExaminer63plot}
\end{figure}

\begin{figure}[H]
	\centering
	\includegraphics[scale=0.4]{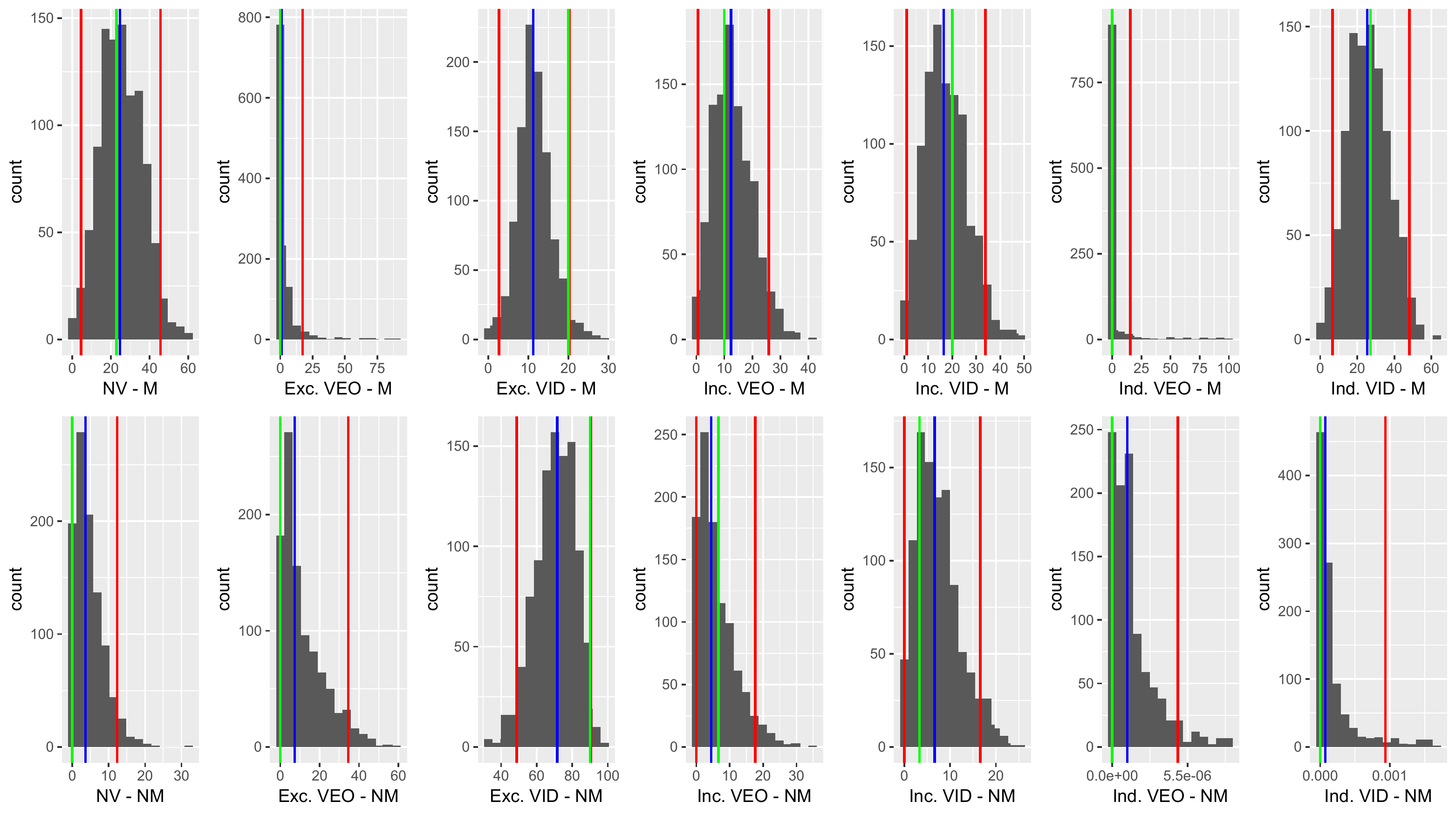}
	\caption{Histograms of the marginal posterior samples of rates (labeled in percent form) from examiner 157. Rates for mated test cases are presented in the top row, while those for non-mated test cases are presented in the bottom row. Vertical blue lines indicate the ABC rate estimates for examiner 157, defined as the median of each marginal posterior sample. The vertical red lines indicate the lower and upper bounds of a 95\% highest posterior density interval for each marginal posterior sample. Vertical green lines indicate the plug-in estimate for the rate of each decision category. Plug-in estimates of the rates for each examiner are obtained in a similar manner as the plug-in estimates for the population rates calculated by \citet{Ulery2011}.}
	\label{fig:ResultsABCExaminer157plot}
\end{figure}

Algorithm \ref{alg:ABCBlackBox} also produces joint posterior samples of the 14 decision rates for each of the 169 examiners in the ``Noblis Black Box" experiment. Based on this data, we define rate estimates for each examiner as the medians of their own marginal posterior samples. Since the posterior samples are drawn from the joint distribution of all 14 rate parameters, the estimates are able incorporate information on the way an examiner makes decisions on all test cases - even across mated and non-mated decisions. Results for examiners 28, 63 and 157 are presented in figures \ref{fig:ResultsABCExaminer28plot} through \ref{fig:ResultsABCExaminer157plot} and tables \ref{tab:ResultsABCExaminer28table} through \ref{tab:ResultsABCExaminer157table}. 

Examiner 28 possesses the highest ABC rate estimate for Ind. VID in the non-mated scenario (i.e., false identification rate for VID fingermarks). However, that examiner only made one error of this type in the experiment, while there was a different examiner who made two errors of this type. Even though examiner 28 did not make the highest number of false identifications in the experiment, this examiner's risk is considered highest. 


In figure \ref{fig:ResultsABCExaminer28plot}, we observe that the ABC estimate for the Ind. VID rate in the non-mated scenario for examiner 28, despite being the highest of all 169 examiners, is quite lower than the corresponding plug-in estimate. The adjustment method described in relation to algorithm \ref{alg:ABCalgorithm} takes into account information from the population rates to adjust each individual's parameters. Hence, the population acts as a regularisation factor in the construction of each individual's posterior sample. 

\begin{table}[H]
\centering
\resizebox{0.75\textwidth}{!}{\begin{tabular}{|l|c|c|c|c|}
  \hline
  \cellcolor{lightgray} & \cellcolor{lightgray}Mated est. & \cellcolor{lightgray}Mated HDI & \cellcolor{lightgray}Non-mated est. & \cellcolor{lightgray}Non-mated HDI \\ \hline
  \cellcolor{lightgray}NV & 26.67\% & [12.31\%, 40.40\%] & 9.24\% & [1.61\%, 17.99\%] \\ \hline
  \cellcolor{lightgray}Exc. VEO & 1.13\% & [0.01\%, 3.64\%] & 3.53\% & [0.00\%, 10.31\%] \\ \hline
  \cellcolor{lightgray}Exc. VID & 2.17\% & [0.39\%, 5.78\%] & 68.15\% & [53.25\%, 83.17\%] \\ \hline
  \cellcolor{lightgray}Inc. VEO & 17.98\% & [7.07\%, 29.33\%] & 10.26\% & [1.53\%, 18.41\%] \\ \hline
  \cellcolor{lightgray}Inc. VID & 18.02\% & [5.67\%, 28.88\%] & 5.04\% & [0.00\%, 12.37\%] \\ \hline
  \cellcolor{lightgray}Ind. VEO & 0.63\% & [0.00\%, 8.37\%] & 0.00\% & [0.00\%, 0.00\%] \\ \hline 
  \cellcolor{lightgray}Ind. VID & 31.15\% & [16.63\%, 45.40\%] & 0.42\% & [0.00\%, 2.43\%] \\
  \hline
\end{tabular}}
\caption{Examiner 28 rate estimates and 95\% highest posterior density intervals for each of the 14 decisions. Values are presented as percents of total mated or non-mated presented pairs and are rounded to two decimal places. Note that since the median are used as point estimates, they do not sum to 100\% across the seven categories for mated comparisons and the seven categories for non-mated comparisons.}
\label{tab:ResultsABCExaminer28table}
\end{table}

Examiner 63 made two erroneous conclusions in the Ind. VID category of the non-mated scenario (i.e., false identifications), yet the ABC rate estimate for that examiner in that decision category is virtually 0. In fact, it is much lower than the corresponding rate in examiner 28, who only committed one error. It may be that this examiner's decision patterns in the other categories indicate a low risk of committing false identifications and that the two observed ones result from (bad) luck. Alternatively, it may be that our data generation algorithm (algorithm \ref{alg:BBdatagen}) does not have the necessary resolution to generate posterior samples for individual examiners. For example, our algorithm does not account for comparison difficulty or examiner competency; and it is constrained by the number of observations collected  by \cite{Ulery2011} for each examiner (i.e., approximately 100).  

\begin{table}[H]
\centering
\resizebox{0.75\textwidth}{!}{\begin{tabular}{|l|c|c|c|c|}
  \hline
  \cellcolor{lightgray} & \cellcolor{lightgray}Mated est. & \cellcolor{lightgray}Mated HDI & \cellcolor{lightgray}Non-mated est. & \cellcolor{lightgray}Non-mated HDI \\ \hline
  \cellcolor{lightgray}NV & 28.59\% & [7.50\%, 48.47\%] & 7.01\% & [0.00\%, 16.58\%] \\ \hline
  \cellcolor{lightgray}Exc. VEO & 1.03\% & [0.00\%, 35.20\%] & 6.21\% & [0.00\%, 16.58\%] \\ \hline
  \cellcolor{lightgray}Exc. VID & 2.65\% & [0.00\%, 21.81\%] & 56.07\% & [40.80\%, 71.95\%] \\ \hline
  \cellcolor{lightgray}Inc. VEO & 22.82\% & [1.93\%, 36.13\%] & 20.33\% & [8.66\%, 31.91\%] \\ \hline
  \cellcolor{lightgray}Inc. VID & 11.88\% & [0.30\%, 23.41\%] & 7.47\% & [0.00\%, 18.60\%] \\ \hline
  \cellcolor{lightgray}Ind. VEO & 0.23\% & [0.00\%, 9.74\%] & 0.00\% & [0.00\%, 0.00\%] \\ \hline 
  \cellcolor{lightgray}Ind. VID & 23.68\% & [1.48\%, 38.44\%] & 0.06\% & [0.00\%, 0.20\%] \\
  \hline
\end{tabular}}
\caption{Examiner 63 rate estimates and 95\% highest posterior density intervals for each of the 14 decisions. Values are presented as percents of total mated or non-mated presented pairs and are rounded to two decimal places. Note that since the median are used as point estimates, they do not sum to 100\% across the seven categories for mated comparisons and the seven categories for non-mated comparisons.}
\label{tab:ResultsABCExaminer63table}
\end{table}

The ABC rate estimate for examiner 157 in the Exc. VID category of the mated scenario (false exclusion) is the highest. This examiner made 14 false exclusions, which is the highest number of false exclusions for VID fingermarks made by any examiner in the \cite{Ulery2011} experiment (see \cite{BBResults} data). The next highest number of false exclusions for VID fingermarks made by a single examiner was 12. We observe that the ABC estimate for Exc. VID for the mated scenario is lower than the plug-in estimate. We believe that we are observing the same regularisation effect from the population data already discussed in relation to examiner 28. Interestingly, we observe that this examiner has a tendency to exclude at a high rate. This is confirmed by the data presented in figure \ref{fig:ExaminerPairsPlotCol3_top20ish} where we note that examiners with the highest rate of false exclusion tend to reach exclusion decisions at a higher rate than the other examiners. 

\begin{table}[H]
\centering
\resizebox{0.75\textwidth}{!}{\begin{tabular}{|l|c|c|c|c|}
  \hline
  \cellcolor{lightgray} & \cellcolor{lightgray}Mated est. & \cellcolor{lightgray}Mated HDI & \cellcolor{lightgray}Non-mated est. & \cellcolor{lightgray}Non-mated HDI \\ \hline
  \cellcolor{lightgray}NV & 24.67\% & [4.60\%, 45.66\%] & 3.64\% & [0.00\%, 12.33\%] \\ \hline
  \cellcolor{lightgray}Exc. VEO & 0.96\% & [0.00\%, 17.27\%] & 7.31\% & [0.00\%, 34.56\%] \\ \hline
  \cellcolor{lightgray}Exc. VID & 11.24\% & [2.68\%, 20.28\%] & 71.51\% & [48.84\%, 90.56\%] \\ \hline
  \cellcolor{lightgray}Inc. VEO & 12.38\% & [0.58\%, 25.83\%] & 4.48\% & [0.00\%, 17.65\%] \\ \hline
  \cellcolor{lightgray}Inc. VID & 16.43\% & [1.05\%, 33.87\%] & 6.65\% & [0.00\%, 16.60\%] \\ \hline
  \cellcolor{lightgray}Ind. VEO & 0.16\% & [0.00\%, 15.49\%] & 0.00\% & [0.00\%, 0.00\%] \\ \hline 
  \cellcolor{lightgray}Ind. VID & 25.32\% & [6.56\%, 48.14\%] & 0.00\% & [0.00\%, 0.00\%] \\
  \hline
\end{tabular}}
\caption{Examiner 157 rate estimates and 95\% highest posterior density intervals for each of the 14 decisions. Values are presented as percents of total mated or non-mated presented pairs and are rounded to two decimal places. Note that since the median are used as point estimates, they do not sum to 100\% across the seven categories for mated comparisons and the seven categories for non-mated comparisons.}
\label{tab:ResultsABCExaminer157table}
\end{table}

The rate estimates and credible intervals for individual examiners presented in tables \ref{tab:ResultsABCExaminer28table} to \ref{tab:ResultsABCExaminer157table} can be compared to their counterpart plug-in estimates and confidence intervals in a similar manner as we did with respect to table \ref{tab:ResultsABCPopulation} for the population. The data, which is not reported here, shows that credible intervals are much wider than the corresponding confidence intervals calculated using the Agresti-Coull method.

\subsection{Potential for predicting individual examiners' risks}
\label{PredictionCluster}

The study of the individual ABC rate estimates may allow to detect behavioural patterns in the decision making of fingerprint examiners. Ultimately, such clustering may be beneficial as it may allow to predict individual examiners' risks based on their observed pattern of rates in casework and their association with one of several predetermined behavioural types. This strategy may allow to identify high risk examiners and take corrective actions before errors occur. 

\begin{figure}[H]
	\centering
	\includegraphics[scale=0.5]{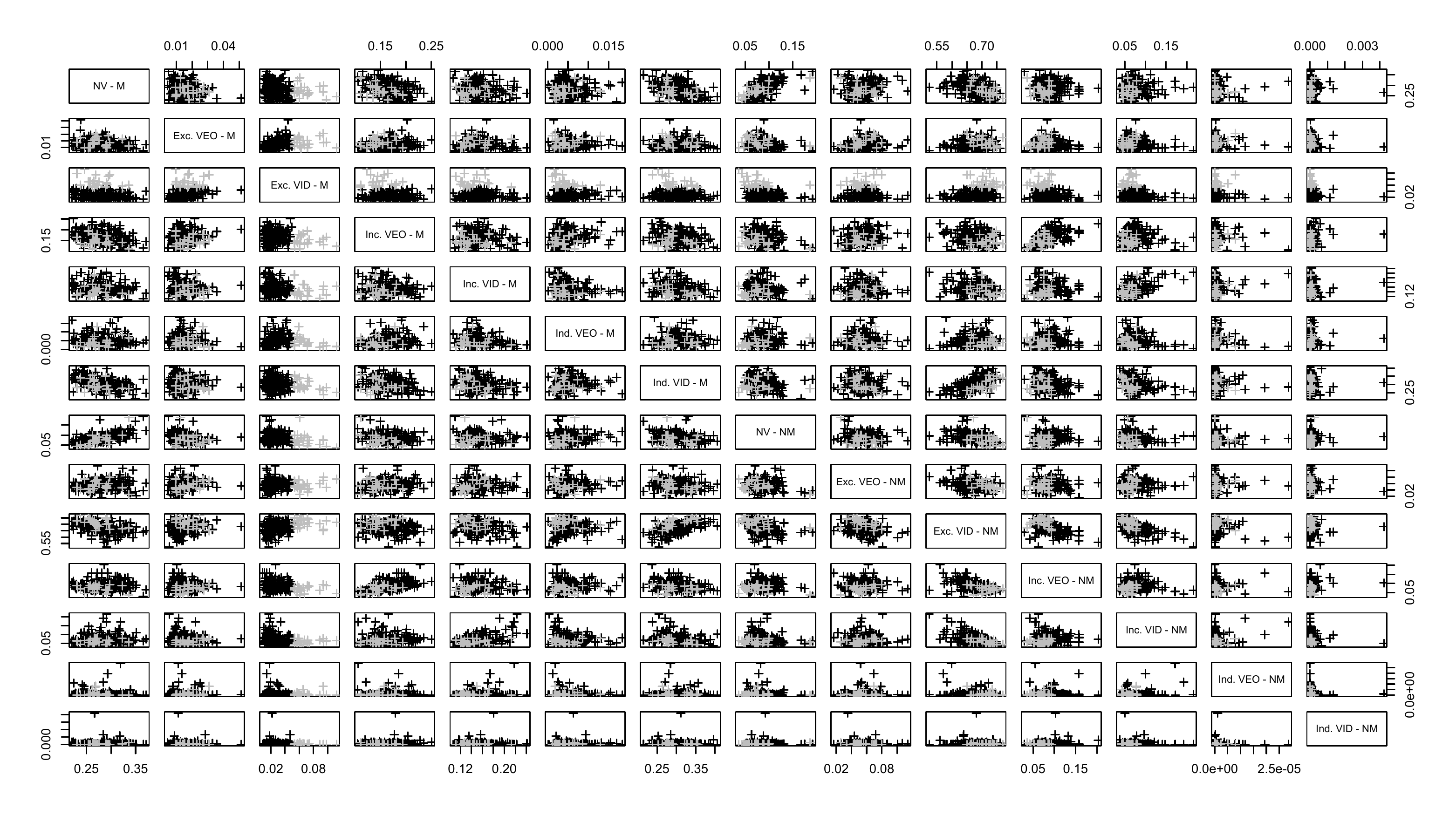}
	\caption{Pairwise plots of the medians for all examiners in each of the 14 decision categories. The data has been partitioned between the examiners with an ABC estimated rate of erroneous exclusion on VID fingermarks larger than 5\% (grey) and the other ones (black).}
	\label{fig:ExaminerPairsPlotCol3_top20ish}
\end{figure}

\begin{figure}[H]
	\centering
	\includegraphics[scale=0.5]{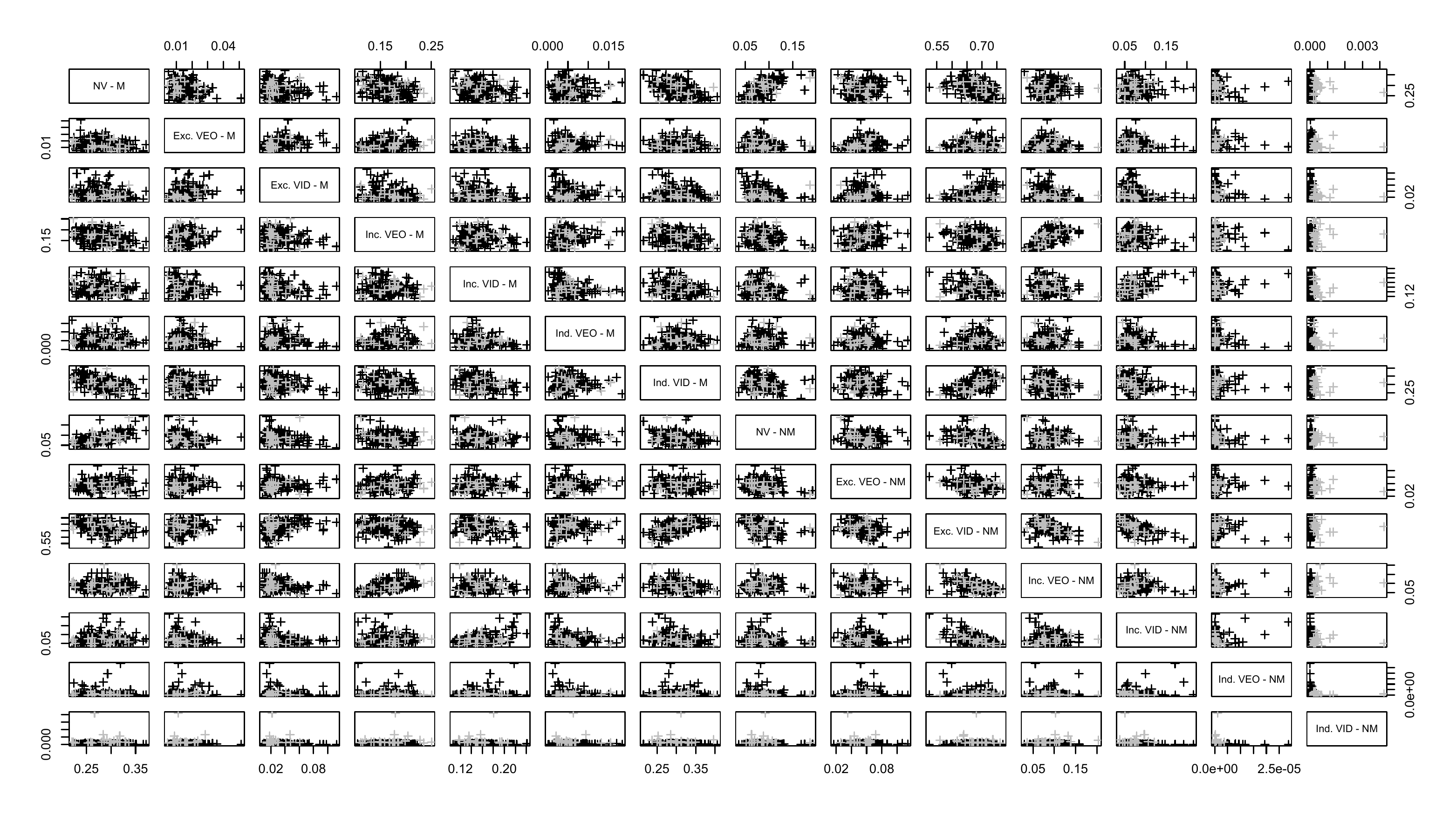}
	\caption{Pairwise plots of the medians for all examiners in each of the 14 decision categories. The data has been partitioned between the examiners with an ABC estimated rate of erroneous identifications on VID fingermarks larger than 0.02\% (grey) and the other ones (black).}
	\label{fig:ExaminerPairsPlotCol14}
\end{figure}

We tried to construct a predictive model based on the \cite{Ulery2011} data; however, our model was not very successful. We believe that we are missing some key information, such as objective measures of image quality and comparison difficulty, as well as agency policies and operating procedures related to VEO and VID determination at the analysis stage of ACE-V. Nevertheless, we believe that such a model is possible if based on appropriate data as indicated by the two examples presented in figures \ref{fig:ExaminerPairsPlotCol3_top20ish} and \ref{fig:ExaminerPairsPlotCol14}. 

In figure \ref{fig:ExaminerPairsPlotCol3_top20ish}, we separate examiners that have ``high" rates of erroneous exclusions on VID fingermarks. We choose to consider examiners whose ABC rate estimates are greater than 5\%. Note that this cut-off threshold is much greater than the median rate observed for the population. We observe that examiners with high rates of false exclusion on VID fingermarks have generally higher rates of exclusion, lower rates of no value decisions, and lower rates of inconclusive decisions. Overall, these examiners seem to take more risk than others (since they are less likely to use the no value and inconclusive categories). They may also be less inclined to identify, but the observations of columns 6, 7, 13 and 14 are not conclusive. 

Figure \ref{fig:ExaminerPairsPlotCol14} enables us to study characteristics of examiners with high rates of erroneous identification on VID fingermarks. We have chosen to consider examiners with ABC rate estimates larger than 0.02\%. Note that this cut-off threshold is very small, but it much greater than the median rate observed for the population. We observe that these examiners have lower rates of mated Exc. VID decisions, lower rates of non-mated Exc. VEO decisions, and lower rates of non-mated Inc. VID decisions.

It is interesting to note that examiners with high risk of false identification for VID fingermarks are not necessarily the examiners with high risk of false exclusion for VID fingermarks (although there is some overlap). 

These two examples make it evident that there are groups of examiners that behave in similar manners, and these could be used to identify individuals that are at risk to commit false exclusions or false identifications. As mentioned above, more advanced clustering methods (than the simple observation of bivariate empirical distributions) could be used to identify these groups in a more formal manner; however, we were unable to do so with the \cite{Ulery2011} study due the relatively small sample size and the unavailability of the key information mentioned previously. 



\section{Conclusion}
\label{Conclusion}

In this paper, we propose an Approximate Bayesian Computation (ABC) method to analyse the results of ``black box'' experiments designed to study the rates of occurrence of erroneous conclusions in feature-based comparison methods. Our method relies on strong statistical foundations and is capable of handling the most complex designs (including unbalanced designs, dependent observations and missing data) in a rigorous manner. Our method also allows for studying the parameters of interest without recourse to incoherent and misleading measures of uncertainty such as confidence intervals. 

To illustrate our method, we used it to reanalyse results of the ``Noblis Black Box" study \citep{Ulery2011}. In general, plug-in estimates for the population rates defined by \cite{Ulery2011} and the corresponding ABC rate estimates agree. Importantly, our method assigns credible intervals to the value of the decision rates. We note that these credible intervals are much wider than the confidence intervals for the corresponding parameter estimates. The Noblis results did not account for the dependencies among observations in the design of the experiment, which resulted in underestimated variances. Our method accounts for these dependencies and reports more reasonable intervals. 

Furthermore, our method allows us to go beyond simply reporting point estimates for decision rates of a population. Our method can quantify the risk of errors of any given examiner, even when no error has been recorded for that examiner. This opens the door to the detection of behavioural patterns in the decision-making of examiners through their ABC rate estimates. These patterns could be used to detect error prone examiners, enabling additional training efforts to be more tailored to each examiner, and allowing for limiting the risk of errors before they even occur. 

We realise that our algorithm \ref{alg:BBdatagen} is not a perfect representation of the Noblis experiment, and therefore, that there is no absolute guarantee that our method can recover the true decision rates of the fingerprint examiner population, or of a given fingerprint examiner; however, the results shown in appendices A and B regarding the adequacy of our data generating process and our ABC algorithm support that our method has the ability to recover the true (but unknown) population and examiner rates that were studied by the ``Noblis Black Box" experiment. 

Finally, all of our efforts were focused on the set of decisions available to examiners over the entire ACE process. Our set of decisions include no value decisions as well as inconclusive. Thus, our rate estimates are not ``normalised'' to only reflect the decisions made on fingermarks deemed of value for comparison/identification. The extension of our work to the sole ``of value'' prints is trivial, but may better uncover patterns in the decision-making of examiners. We have requested the release of the data originating from the ``Noblis White Box'' experiment and we plan to extend our work on the predictive model using this new dataset when available.


\bibliographystyle{chicago}
\bibliography{ABC_for_error_rates}

\section*{Appendix A}
\label{AppendixA}

One of the necessary assumptions for the method of analysis that we propose above is that the data generating algorithm appropriately simulates the considered experiment. We wish to verify the robustness of this assumption with respect to algorithm \ref{alg:BBdatagen} and the ``Noblis Black Box" experiment described by \citet{Ulery2011}. 

Our strategy is to simulate the Noblis experiment a large number of times using the same values for the decision rates in each run of the simulations. We choose to use the point estimates for the decision rates of each examiner calculated from the observed data acquired by \cite{Ulery2011}. This choice seems reasonable to best describe the participants to the Noblis experiment. 

We define the inputs to the algorithm as follows:
\spacing{1}
\begin{enumerate}
	\item $n_{m}^{(j)}$ and $n_{nm}^{(j)}$, for $j = 1,2, ..., 169$, using the true values from the Noblis experiment; 
	\item $\theta_{m}^{(j)}$ and $\theta_{nm}^{(j)}$, for $j = 1,2, ..., 169$, using the point estimates for each examiner separately obtained from the Noblis experiment. 
\end{enumerate}	
\spacing{1.5}

We tallied the total number of pseudo-decisions observed in each category for each run of the simulations. In table \ref{tab:ResultsValidationAlg3}, we present the lower and upper bounds for the 95\% highest density intervals for each decision category. 

\begin{table}[h]
\centering
\begin{tabular}{|l|c|c|c|c|}
  \hline
  \cellcolor{lightgray} & \cellcolor{lightgray}Mated & \cellcolor{lightgray}Mated & \cellcolor{lightgray}Non-mated & \cellcolor{lightgray}Non-mated \\
  \cellcolor{lightgray} & \cellcolor{lightgray}observed counts & \cellcolor{lightgray}simulated HDI & \cellcolor{lightgray}observed counts & \cellcolor{lightgray}simulated HDI \\ \hline
  \cellcolor{lightgray}NV & 3389 & [3296,3482] & 558 & [515,599] \\ \hline
  \cellcolor{lightgray}Exc. VEO & 161 & [135,184] & 325 & [291,356] \\ \hline
  \cellcolor{lightgray}Exc. VID & 450 & [409,488] & 3622 & [3555,3687] \\ \hline
  \cellcolor{lightgray}Inc. VEO & 2019 & [1941,2097] & 577 & [533,617] \\ \hline
  \cellcolor{lightgray}Inc. VID & 1856 & [1778,1928] & 455 & [416,491] \\ \hline
  \cellcolor{lightgray}Ind. VEO & 40 & [27,51] & 0 & [0,0] \\ \hline
  \cellcolor{lightgray}Ind. VID & 3663 & [3570,3764] & 6 & [1,10] \\
  \hline
\end{tabular}
\caption{Observed counts in the Noblis experiment for mated and non-mated comparisons and lower and upper bounds for the 95\% highest density interval resulting from 10,000 simulations using algorithm \ref{alg:BBdatagen}.}
\label{tab:ResultsValidationAlg3}
\end{table}

Table \ref{tab:ResultsValidationAlg3} shows that the true observed counts from the Noblis experiment fall within the simulated ranges. Therefore, we consider that our data generating algorithm appropriately reproduces the Noblis experiment.

\section*{Appendix B}
\label{AppendixB}

We also wish to verify that our ABC algorithm (algorithm \ref{alg:ABCBlackBox}) has the ability to recover the true population rates and the true examiner rates underlying the ``Noblis Black Box" experiment. Since the true rates are unknown, we develop a test scenario in which we define a set of ``true" population rates. ``True" individual examiner rates are randomly sampled from Dirichlet distributions with parameters based on the ``true" population rates. ``True" rates for the population and several individual examiners are presented in table \ref{tab:TrueRates}.

\begin{table}[h]
\centering
\resizebox{\textwidth}{!}{\begin{tabular}{|l|c|c|c|c|c|c|c|c|}
  \hline
  \cellcolor{lightgray} & \multicolumn{2}{|c|}{\cellcolor{lightgray} Population} & \multicolumn{2}{|c|}{\cellcolor{lightgray}Examiner 10} & \multicolumn{2}{|c|}{\cellcolor{lightgray}Examiner 42} & \multicolumn{2}{|c|}{\cellcolor{lightgray}Examiner 107} \\
  \cellcolor{lightgray} & \multicolumn{1}{|c}{\cellcolor{lightgray}Mated} & \multicolumn{1}{c|}{\cellcolor{lightgray}Non-mated} & \multicolumn{1}{|c}{\cellcolor{lightgray}Mated} & \multicolumn{1}{c|}{\cellcolor{lightgray}Non-mated} & \multicolumn{1}{|c}{\cellcolor{lightgray}Mated} & \multicolumn{1}{c|}{\cellcolor{lightgray}Non-mated} & \multicolumn{1}{|c}{\cellcolor{lightgray}Mated} & \multicolumn{1}{c|}{\cellcolor{lightgray}Non-mated} \\
  \hline
  \cellcolor{lightgray}NV & \multicolumn{1}{|r|}{30.70\%} & \multicolumn{1}{r|}{10.90\%} & \multicolumn{1}{|r|}{32.36\%} & \multicolumn{1}{r|}{14.15\%} & \multicolumn{1}{|r|}{33.74\%} & \multicolumn{1}{r|}{26.74\%} & \multicolumn{1}{|r|}{35.17\%} & \multicolumn{1}{r|}{19.72\%} \\ \hline
  \cellcolor{lightgray}Exc. VEO & \multicolumn{1}{|r|}{1.00\%} & \multicolumn{1}{r|}{6.00\%} & \multicolumn{1}{|r|}{3.02\%} & \multicolumn{1}{r|}{4.01\%} & \multicolumn{1}{|r|}{1.11\%} & \multicolumn{1}{r|}{7.17\%} & \multicolumn{1}{|r|}{0.00\%} & \multicolumn{1}{r|}{11.57\%} \\ \hline
  \cellcolor{lightgray}Exc. VID & \multicolumn{1}{|r|}{4.00\%} & \multicolumn{1}{r|}{65.00\%} & \multicolumn{1}{|r|}{0.82\%} & \multicolumn{1}{r|}{79.18\%} & \multicolumn{1}{|r|}{0.07\%} & \multicolumn{1}{r|}{46.08\%} & \multicolumn{1}{|r|}{0.52\%} & \multicolumn{1}{r|}{57.87\%} \\ \hline
  \cellcolor{lightgray}Inc. VEO & \multicolumn{1}{|r|}{17.00\%} & \multicolumn{1}{r|}{10.00\%} & \multicolumn{1}{|r|}{14.35\%} & \multicolumn{1}{r|}{1.46\%} & \multicolumn{1}{|r|}{25.15\%} & \multicolumn{1}{r|}{5.14\%} & \multicolumn{1}{|r|}{17.19\%} & \multicolumn{1}{r|}{10.51\%} \\ \hline
  \cellcolor{lightgray}Inc. VID & \multicolumn{1}{|r|}{16.00\%} & \multicolumn{1}{r|}{8.00\%} & \multicolumn{1}{|r|}{3.52\%} & \multicolumn{1}{r|}{1.20\%} & \multicolumn{1}{|r|}{3.81\%} & \multicolumn{1}{r|}{14.87\%} & \multicolumn{1}{|r|}{15.98\%} & \multicolumn{1}{r|}{0.25\%} \\ \hline
  \cellcolor{lightgray}Ind. VEO & \multicolumn{1}{|r|}{0.30\%} & \multicolumn{1}{r|}{0.00\%} & \multicolumn{1}{|r|}{4.34\%} & \multicolumn{1}{r|}{0.00\%} & \multicolumn{1}{|r|}{0.08\%} & \multicolumn{1}{r|}{0.00\%} & \multicolumn{1}{|r|}{0.00\%} & \multicolumn{1}{r|}{0.00\%} \\ \hline
  \cellcolor{lightgray}Ind. VID & \multicolumn{1}{|r|}{31.00\%} & \multicolumn{1}{r|}{0.10\%} & \multicolumn{1}{|r|}{41.60\%} & \multicolumn{1}{r|}{0.00\%} & \multicolumn{1}{|r|}{36.04\%} & \multicolumn{1}{r|}{0.00\%} & \multicolumn{1}{|r|}{31.14\%} & \multicolumn{1}{r|}{0.08\%} \\ \hline       
\end{tabular}}
\caption{``True" population rates and ``true" rates for examiners 10, 42, and 107, as defined for the simulation experiment. Rates are presented as percents, rounded to two decimal places.}
\label{tab:TrueRates}
\end{table}

Using algorithm \ref{alg:BBdatagen} and this set of ``true" rates, we simulate a vector of counts representing the total number of decisions in each of the 14 decision categories. Using algorithm \ref{alg:ABCBlackBox}, we then try to recover the ``true'' rates. The inputs for algorithms \ref{alg:BBdatagen} and \ref{alg:ABCBlackBox} are defined as follows:
\spacing{1}
\begin{enumerate}
	\item The number of simulations, $N$, is 100,000;
	\item The parameters of the Dirichlet distribution for the decision rates of the population for mated comparisons are $\alpha_{m}=[3, 0.5, 0.5, 2, 2, 0.5, 3]$.
	\item The parameters of the Dirichlet distribution for the decision rates of the population for non-mated comparisons are $\alpha_{nm}=[1, 1, 7, 1, 1, 0.5, 0.5]$.
	\item The parameters for the lognormal distribution for a scale parameter used in sampling of the decision rates for the examiners from the population rates are $\mu=6$ and $\sigma=1$.
	\item The numbers of mated and non-mated image pairs $n^{(j)}_{m}$ and $n^{(j)}_{nm}$ presented to examiners $j=1,...169$ are obtained from the observed counts for each examiner in the ``Noblis Black Box" study.
	\item The observed data, $D^*_{obs}$, is the set of simulated decisions produced using the set of defined ``true" rates.
\end{enumerate}
\spacing{1.5}

Note that the posterior samples resulting from the use of algorithm \ref{alg:ABCBlackBox} are produced without knowledge of the ``true" population and examiner rates. 

\begin{figure}[h]
	\centering
	\includegraphics[scale=0.4]{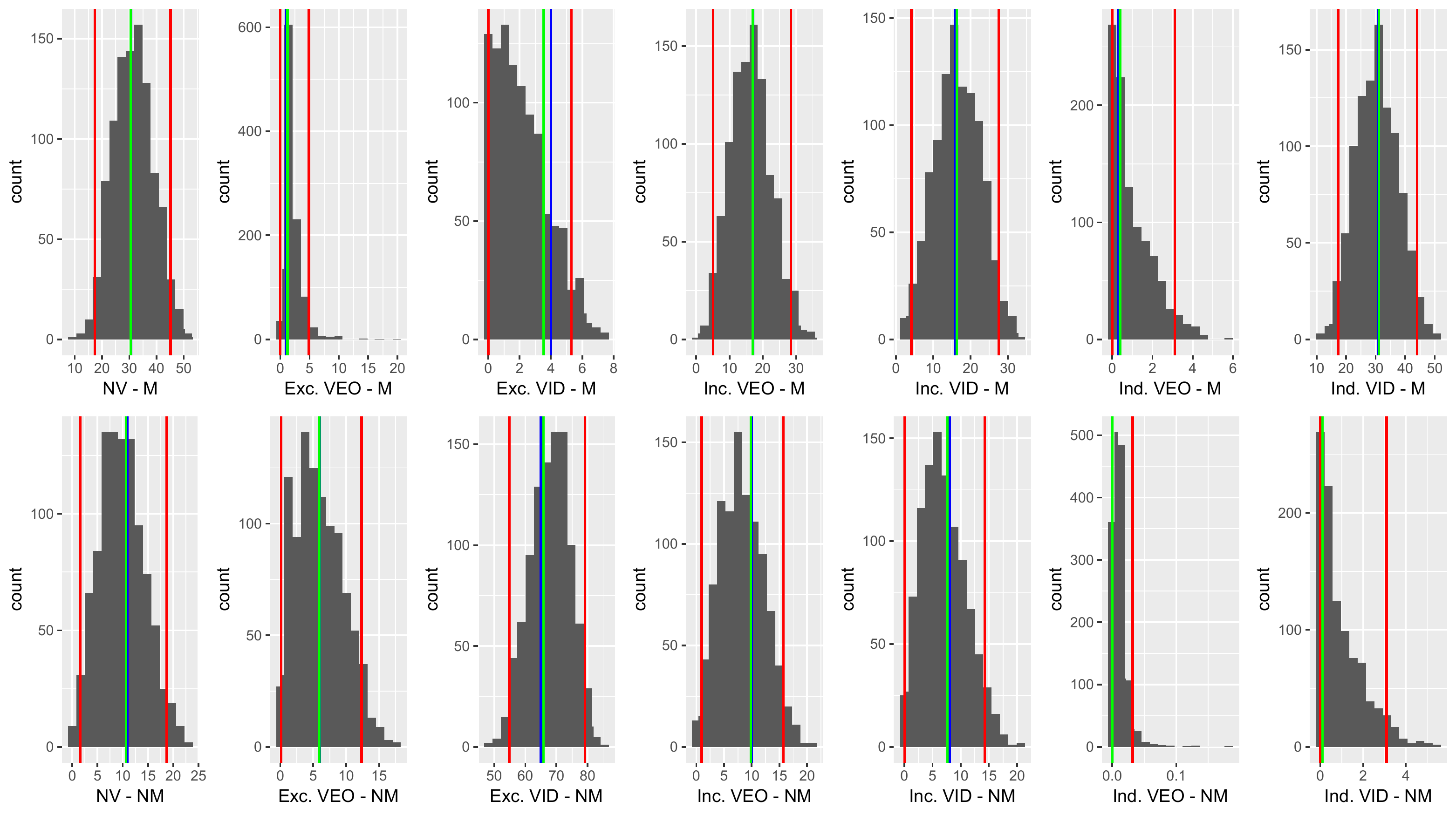}
	\caption{Results of the ABC verification experiment. Histograms of the marginal posterior samples of rates for mated test cases are presented in the top row, while those for non-mated rates are presented in the bottom row. Vertical blue lines indicate the ``true" population rates defined for this experiment. Vertical green lines indicate the plug-in estimates for the rates of each decision category based on the decision counts, $D^*_{obs}$, obtained from the initial run of algorithm \ref{alg:BBdatagen}. The vertical red lines indicate the lower and upper bounds of the 95\% highest posterior density interval for each marginal posterior sample.}
	\label{fig:VerifyABCPopulation}
\end{figure}

\begin{figure}[h]
	\centering
	\includegraphics[scale=0.4]{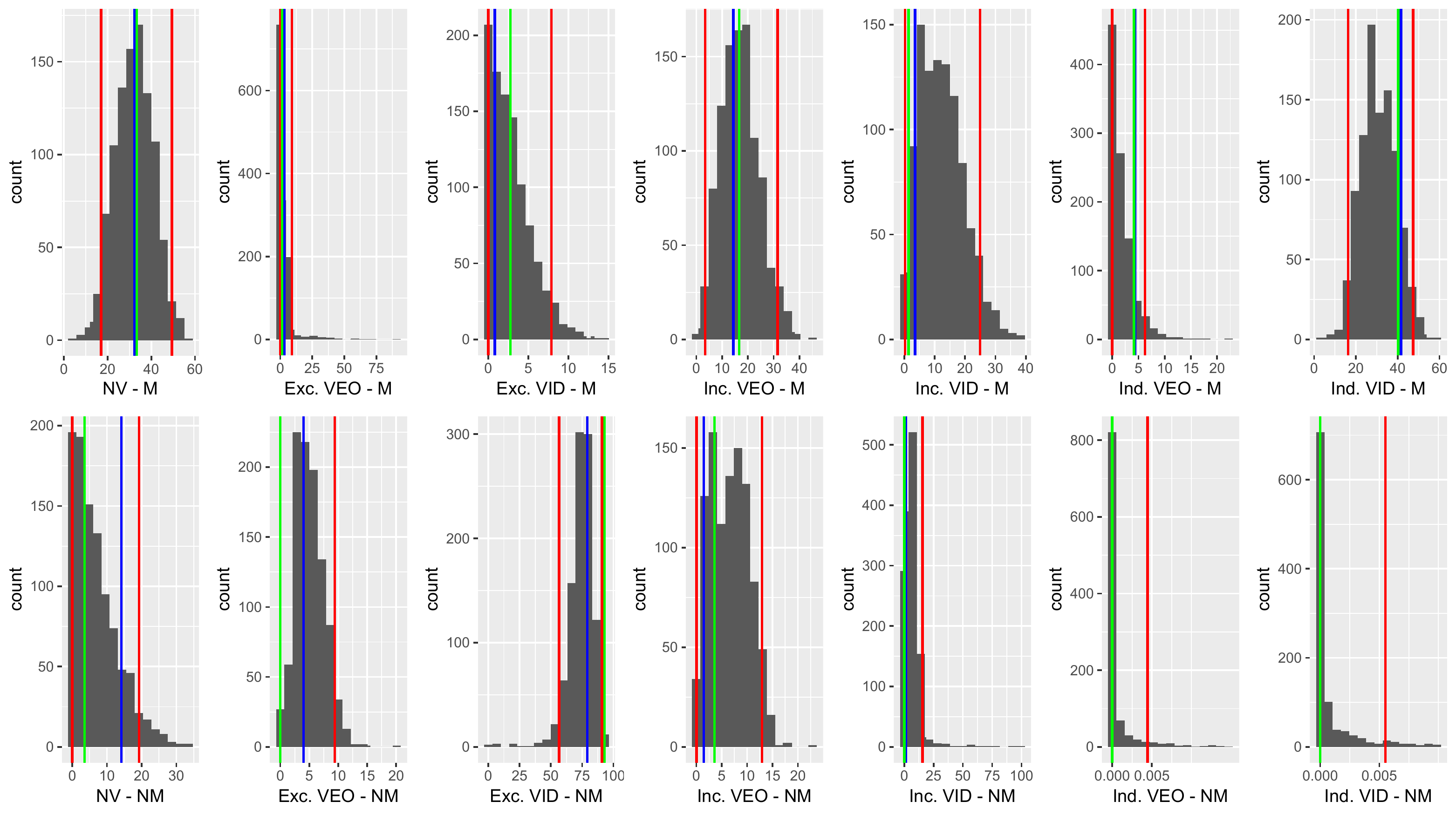}
	\caption{Results of the ABC verification experiment for examiner 10. Histograms of the marginal posterior samples of rates for mated test cases are presented in the top row, while those for non-mated rates are presented in the bottom row. Vertical blue lines indicate the ``true" rates defined for examiner 10 in this experiment. Vertical green lines indicate the plug-in estimates for the rates of each decision category based on the decision counts, $D^*_{obs}$, obtained from the initial run of algorithm \ref{alg:BBdatagen}. The vertical red lines indicate the lower and upper bounds of the 95\% highest posterior density interval for each marginal posterior sample.}
	\label{fig:VerifyABCExaminer10}
\end{figure}

\begin{figure}[h]
	\centering
	\includegraphics[scale=0.4]{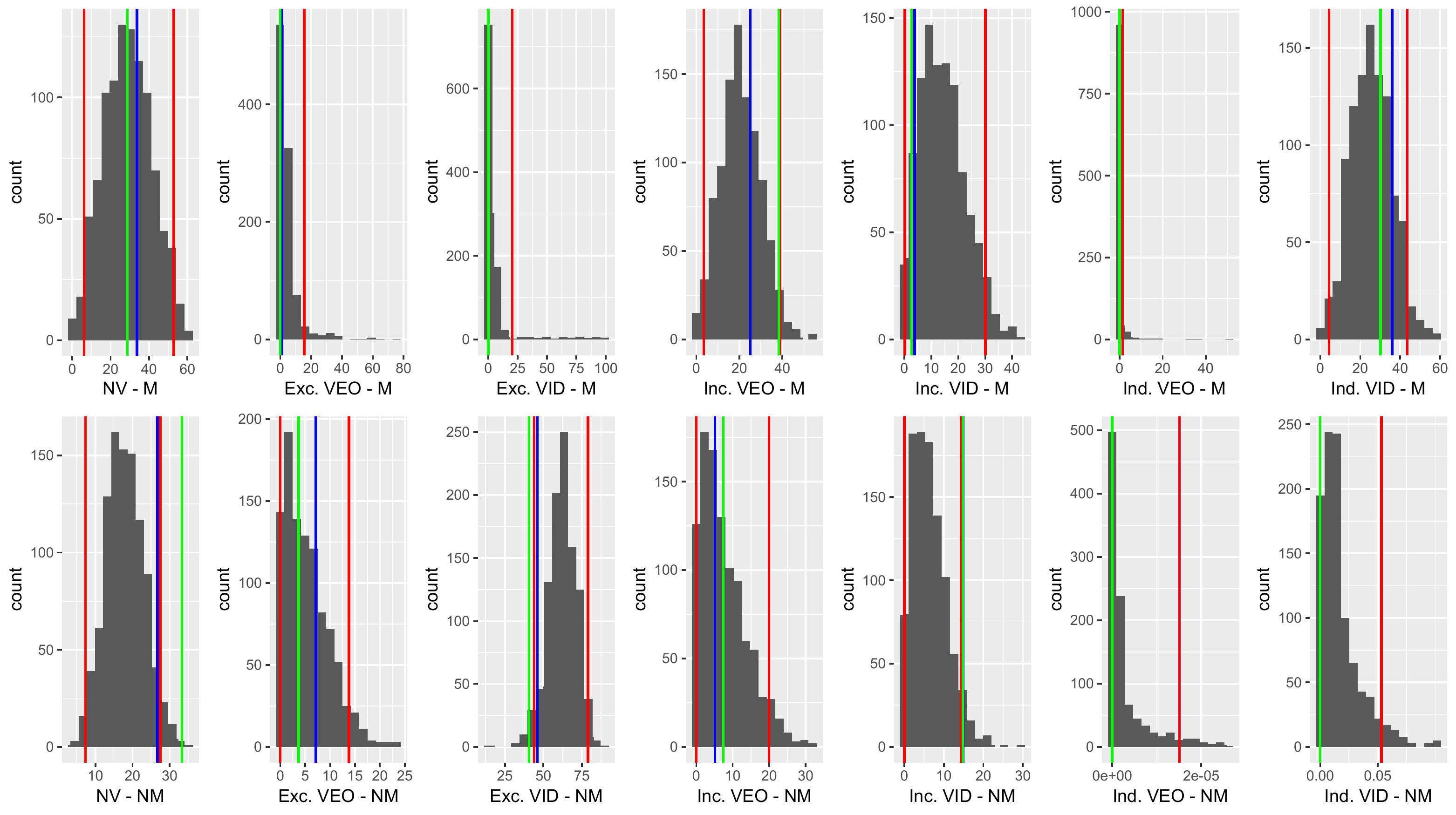}
	\caption{Results of the ABC verification experiment for examiner 42. Histograms of the marginal posterior samples of rates for mated test cases are presented in the top row, while those for non-mated rates are presented in the bottom row. Vertical blue lines indicate the ``true" rates defined for examiner 42 in this experiment. Vertical green lines indicate the plug-in estimates for the rates of each decision category based on the decision counts, $D^*_{obs}$, obtained from the initial run of algorithm \ref{alg:BBdatagen}. The vertical red lines indicate the lower and upper bounds of the 95\% highest posterior density interval for each marginal posterior sample.}
	\label{fig:VerifyABCExaminer42}
\end{figure}

\begin{figure}[h]
	\centering
	\includegraphics[scale=0.4]{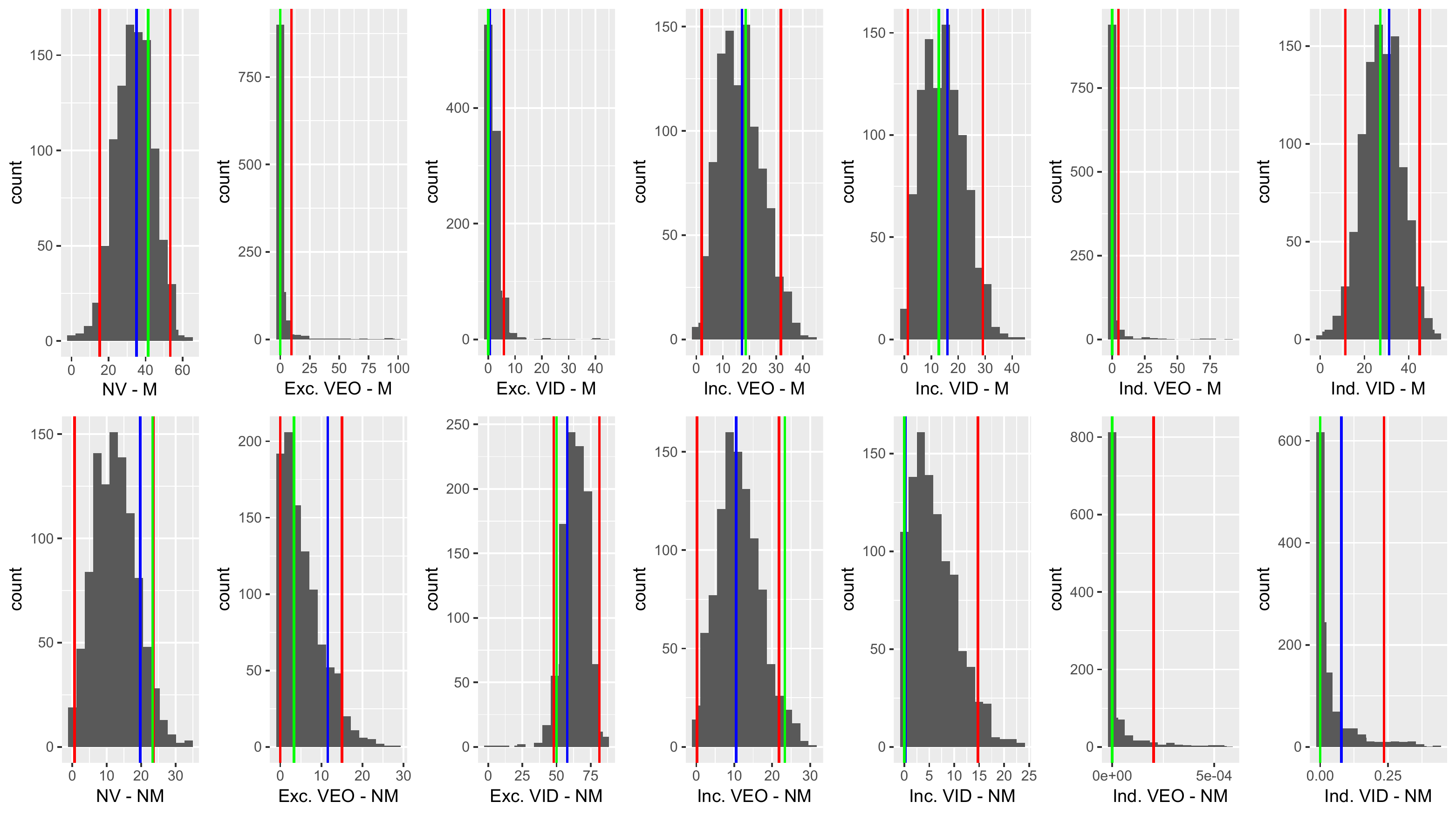}
	\caption{Results of the ABC verification experiment for examiner 107. Histograms of the marginal posterior samples of rates for mated test cases are presented in the top row, while those for non-mated rates are presented in the bottom row. Vertical blue lines indicate the ``true" rates defined for examiner 107 in this experiment. Vertical green lines indicate the plug-in estimates for the rates of each decision category based on the decision counts, $D^*_{obs}$, obtained from the initial run of algorithm \ref{alg:BBdatagen}. The vertical red lines indicate the lower and upper bounds of the 95\% highest posterior density interval for each marginal posterior sample.}
	\label{fig:VerifyABCExaminer107}
\end{figure}

The marginal posterior samples for the population rates are presented in figure~\ref{fig:VerifyABCPopulation}. The marginal posterior samples for the three examiners listed in table \ref{tab:TrueRates} are presented in figures~\ref{fig:VerifyABCExaminer10} to \ref{fig:VerifyABCExaminer107}. We note that the highest density intervals based on the marginal posterior samples contain the ``true" rates in all categories. That said, we have observed that, for some examiners, some of the credible intervals do not include the true value of the parameter. Algorithm \ref{alg:ABCBlackBox} uses the population information to stabilise individual examiner's parameters adjustment. This process has benefits (it ensures that the maximum information is accounted for when adjusting the parameters) but it also has limitations. For example, the dimension of the summary statistics is doubled, which increases the effect of the curse of dimensionality. Furthermore, using population counts as part of the adjustment process may ``steer'' individual examiner's parameter adjustment towards the population estimates.  

 We realise that our algorithm \ref{alg:BBdatagen} is not a perfect representation of the Noblis experiment, and therefore, that there is no absolute guarantee that our method can recover the true decision rates of the fingerprint examiner population studied by \cite{Ulery2011}, or of a given fingerprint examiner; however, the results shown in appendix A regarding the adequacy of our data generating process and the results presented in figures \ref{fig:VerifyABCPopulation} to \ref{fig:VerifyABCExaminer107} support that our method has the ability to recover the true (but unknown) population and examiner rates that were studied by the ``Noblis Black Box" experiment. 

\end{document}